# Transport-domain applications of widely used data sources in the smart transportation: A survey

## Sina Dabiri [a,*], Kevin Heaslip [b]

[a] Charles E. Via, Jr. Department of Civil and Environmental Engineering, 301, Patton Hall, Virginia Tech, Blacksburg, VA 24061, United States
[*] Corresponding Author
Email Addresses: sina@vt.edu, kheaslip@vt.edu

## Abstract

The rapid growth of population and the permanent increase in the number of vehicles engender several issues in transportation systems, which in turn call for an intelligent and cost-effective approach to resolve the problems in an efficient manner. Smart transportation is a framework that leverages the power of Information and Communication Technology for acquisition, management, and mining of traffic-related data sources, which, in this study, are categorized into: 1) traffic flow sensors, 2) video image processors, 3) probe people and vehicles based on Global Positioning Systems (GPS), mobile phone cellular networks, and Bluetooth, 4) location-based social networks, 5) transit data with the focus on smart cards, and 6) environmental data. For each data source, first, the operational mechanism of the technology for capturing the data is succinctly demonstrated. Secondly, as the most salient feature of this study, the transport-domain applications of each data source that have been conducted by the previous studies are reviewed and classified into the main groups. Thirdly, a number of possible future research directions are provided for all types of data sources. Moreover, in order to alleviate the shortcomings pertaining to each single data source and acquire a better understanding of mobility behavior in transportation systems, the data fusion architectures are introduced to fuse the knowledge learned from a set of heterogeneous but complementary data sources. Finally, we briefly mention the current challenges and their corresponding solutions in the smart transportation.

# 1. Introduction

A land transportation network is any platform that permits people to move from an origin to a destination with various modes such as car, bus, train, walking, bicycle, etc. Despite the mobility and access benefits of transportation networks, cities have always been struggling with transportation-related issues (e.g., traffic congestion and air pollution) due to the ever-growing increase in the population and the number of vehicles. Smart transportation, as a cost-effective and significant component of the smart city, has the promising goal to relieve issues and improve cities' livability, workability, and sustainability by developing intelligent and novel transport models. The ultimate efficacy of smart transportation architectures is highly contingent on the quality of deployed data and characteristics of the analytical techniques into which the data are fed. In order to investigate the available traffic-related data sources, we introduce the widely used technologies that are either designed specifically for traffic purposes or potential to be thought of as a traffic data source. These technologies include traffic flow sensors, video image processors, Global Positioning Systems, mobile phone cellular networks, Bluetooth, social network media, smart cards, automated passenger counters, and permanent monitoring stations for collecting environmental data.

Accordingly, the specific objectives of this study are to, first, briefly describe the operational mechanism of the above-mentioned technologies for generating traffic-related data. Secondly, for each data source, we systematically review the transport-domain applications, aiming to classify the past studies into the main groups. For the sake of feasibility, we concentrate on only representative examples in each of those main groups rather than attempt to exhaustively cover all existing models. Thirdly, with regard to each data source, a number of possible fruit areas for future research are provided. We also present data fusion architectures, which provide the opportunity to fuse the knowledge learned from the above-mentioned heterogeneous data sources so as to come up with more improved control systems that ensure safety and an efficient traffic movement.

In this study, we first define the smart city and its general framework in Section 2. Then, we introduce the smart transportation, as one of the main functions in the smart city. After introducing the concepts of the smart city and the smart transportation, the remainder of this article focuses on the data sources that have broadly been utilized in literature to make transportation systems more intelligent. In Sections 3-8, for each above-mentioned data source, we set out their operational mechanisms, transportation-related applications, and future research directions. *It is worth noticing that an interested reader in only one of these data sources can simply skip other sections without losing much information and coherence of this paper*. In Section 9, we introduce the data fusion architectures. Finally, we discuss the current issues in the smart transportation and conclude the paper in Sections 10 and 11, respectively.

# 2. Smart city and smart transportation

## 2.1. Smart city

The rapid growth of various components of a city, including e-government and IT projects, technology, governance, policy, people and community, built infrastructure, and the natural environment, has created a huge complex system (1). Such a complicated system brings about a variety of challenges and risks, with examples ranging from air pollution and traffic congestion, to an increase in the unemployment rate and adverse social effects. Making the cities

"smart", using Information and Communication Technology (ICT), is one solution to manage the urban troubles and enhance cities' livability, workability, and sustainability (2). Although the concept of the smart city is not novel, academics from different fields have defined this term in ways that are not necessarily consistent. For example, smartness in the marketing language focuses on users' perspectives, whereas the smart concept in the urban planning field is defined as new strategies for improving the quality of life and having the sustainable environment (3).

Notwithstanding that many definitions for "smart city" exist in literature, none of them has universally been acknowledged in academia and industry. The most cited and well-known examples of the smart city definition have been provided (3, 4). Much of the smart city research has highlighted one or a number of its aspects including human quality of life, intelligent computing technologies, environmental impacts, condition assessment of infrastructures, social inclusion, and accessibility to the public and private services. Nonetheless, a concrete definition, which is able to transfer the goal of a large portion of amplitude definitions, is: a fusion of ideas about how ICT infrastructures, in addition to human and environment resources, might enhance the quality of life, make the city's infrastructures more intelligent, develop economy and mobility, as well as present efficient methods to address social deprivation and environmental problems (5, 6).

A general framework of the smart city contains three layers: 1) data collection and management, 2) data analytics, and 3) service providing (7). In building a smart city, first, data need to be collected with the aid of smart devices located throughout the city such as traffic sensors, utility usage sensors, weather stations, mobile phones, and social media networks. The heterogeneous data from various sources must be organized using data preprocessing and data management techniques to prepare spatiotemporal information as the main inputs for the data analytics step. After collecting and providing the clean and validated data, data-mining techniques are employed to extract useful knowledge that allows people and machines to act and decide more intelligently. Major comprehension of the current system performance by visualizing and summarizing the associated data, optimizing the complex systems such as coordinating traffic signals, predicting an occurrence of natural disasters, and detecting an anomaly in human mobility are only a few achievements yielded by rigorously mining the sensed data. In the service providing step, authorities are well informed of the current shortcomings of systems and the efficient strategies for outperforming the systems. Maintaining the structural health of buildings, adopting proper priorities in energy consumption behaviors, issuing laws for reducing noises in specific locations, find healthiest paths for outdoor activities, and providing public transits are the services that can be catered for boosting the workability and sustainability of a city (8). Furthermore, predicting the outcome of possible scenarios for using a system (e.g., the amount of fuel consumption and total travel time of choosing different routes when traveling to a destination) enable users to select the appropriate option according to their needs.

As shown in Fig. 1, the smart city includes multiple functions such as smart economy, smart governance, smart people, smart transportation and mobility, smart environment, and smart living (6). Notwithstanding of a strong interconnection between these functions, this survey takes into account only the "smart transportation". It should be noted that the three steps in the smart city architecture is applied to the smart transportation as well.

## 2.2. Smart transportation

Today, every out-of-home activity relies on the transportation system that comprises streets, railways, subways, traffic signals, vehicles, bicycles, buses, etc. A system that moves people around the city for commuting to work, shopping, traveling, going to schools, and hanging out with friends. On the other side, the prompt evolution of transportation networks in cities around the world gives rise to significant troubles. The principal challenge is traffic congestion due to a dramatic increase in transportation modes and population. Besides the direct adverse effects of traffic congestion on users, such as longer trip time and road rage behaviors, it also has long-term negative impacts on energy consumption and air quality (9), economic growth (10), public health (11), and traffic accidents (12-14).

Various attitudes exist to resolve the above and other transportation-related issues (15). One way is to enforce laws on users. For example, preventing the private vehicle owners from entering in the central business district from morning until late afternoon leads to reducing congestion and forcing people to use the public transit. Augmenting the system capacity by constructing new infrastructures, such as bridges, tunnels, and increasing the number of lanes in expressways is another solution for reducing the congestion. However, a cost-effective approach is to optimize the transportation network by extracting hidden knowledge in the data collected from various sources, such as mobile phone networks, commuting smart cards, as well as traffic sensors. The latter approach for tackling with the transport issues constitutes the core of smart transportation or intelligent transportation systems (ITS). Quality and quantity, processing and fusing, and analyzing and discovering new patterns in the collected data are the underlying steps to converting raw data into helpful information applicable in all ITS functional areas, including advanced transportation management systems, advanced travel information systems, advanced vehicle control systems, commercial vehicle operations, advanced public transportation systems, and advanced rural transportation systems (16).

Smart transportation, as one of the smart city functions, is also pursuing the main goals of smart cities, which are ameliorating a city's livability, workability, and sustainability (2). In the livability context, which seeks for a better quality of life for citizens, the use of ICT and advanced traffic data analytics leads to trip time reduction, coordinated traffic signal controls, in-vehicle collision-avoidance systems, ridesharing apps, and public safety. In the workability aspects, which means having better jobs and economic status, business appointments and collaborations between industry owners can be achieved through a reliable transportation network. Moreover, business owners intend to invest in locations with high mobility accessibility. Sustainability is accomplished by efficiently using natural, human, and financial sources while they are not entirely used up or ruined permanently. In pursuit of fulfilling sustainable sources, smart technologies and knowledge discovery can be deployed to reduce air and noise pollution by means of efficient technologies such as route navigation systems. Furthermore, properly assigning budgets for expanding transit services in areas with higher transportation demand results in not only making balance between demand and supply but minimizing energy consumption and pollutant emissions.

Keeping in view the significance of having effective data acquisition and management systems as well as high-quality data analytic models in the smart transportation, in this paper, we investigate the current and widely data sources that various studies have used to come up with novel ideas and address the issues pertaining to transportation systems. A data source, in this study, is only referred to the technologies and systems that have potential to permanently collect traffic data. Data measured in the laboratories or collected by hiring subjects for a specific experiment are not within the scope of this survey. Accordingly, we are seeking to answer the following questions: What are the potential data sources in the smart transportation? What type of technologies and methodologies are utilized to extract and collect transport-domain data from each traffic data source? What are the transportation-related applications associated with each traffic data source?

In pursuit of answering the first question, we categorize traffic data sources into six groups: 1) traffic flow sensors data, 2) video image processors (VIP) data, 3) probe vehicles and people data, 4) location-based social networks data, 5) transit data with the focus on smart cards data, and 6) environmental data. Due to rapid advances in computer vision and image processing fields, VIP has brought much more traffic applications compared to other traffic flow sensors. So we consider VIP as a separate group although it belongs to the traffic flow sensor group. The probe vehicle and people group, which refers to those moving sensors that receive spatiotemporal traffic data, are subdivided based on the technology used to collect the data: Global Positioning Systems (GPS), mobile phone cellular networks, and Bluetooth. Fig. 1 illustrates the structure of traffic data source categorization.

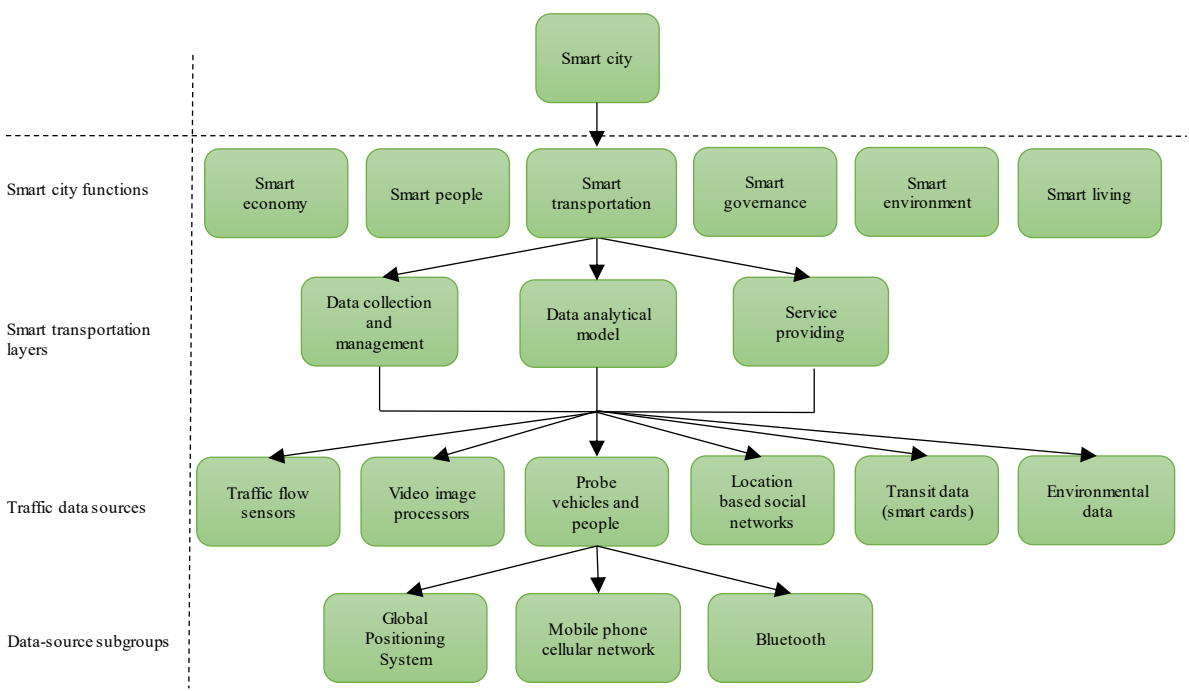

Fig. 1. Smart city and smart transportation architecture with traffic-related data sources.

## 3. Traffic flow sensors

According to Traffic Detector Handbook (17), a traffic flow sensor is a device for capturing the presence or passage of vehicles in a particular location so as to determine traffic states and parameters. The sophisticated data supplied by traffic flow sensors should support important initiatives in all functions of ITS to maximize capacity, alleviate congestion and delay, and improve the reliability of transportation systems. Traffic sensors, which all are considered as point sensors, are categorized into two families: 1) in-roadway sensors that are embedded in pavement or attached to road surfaces, including inductive-loop detectors and magnetometers sensors, 2) over-roadway sensors that are mounted above the surface, including VIP, microwave radar sensors, laser radar sensors, and ultrasonic/acoustic/passive infrared sensors. Detection systems can be built up based on one of the traffic flow sensors or a combination of them. Installation and maintenance cost, traffic disruption and safety of installers, coverage area with multiple lanes and detection zones, easiness of working, providing high quality data with respect to weather effects and variable lightening, traffic flow conditions and the number of measured parameters are the criteria that should be taken into account for opting the appropriate detection system (17).

Both types of sensor technologies bring strengths and weaknesses. In contrast to over-roadway sensors that are affected by weather-related issues, the in-roadway sensors are insensitive to weather conditions due to their close location to the vehicle (17). However, installation and maintenance of in-roadway sensors require road closure and physical changes on the road surface. The main advantage of over-roadway sensors compared to the in-roadway technologies is their capability to monitor multiple lanes or create multi-detection zones in one lane at the same time with only one unit, whereas multiple loops and magnetometers are required to screen all lanes of an approach.

This section begins with describing the operational mechanisms of the sensors. It will then go to the general review of applications of fixed-point detection systems in traffic monitoring and control strategies. The general outline of traffic flow sensors and their associated applications is shown in Fig. 2. Since the information provided by various traffic sensors is usually similar to each other, their associated applications are reviewed under the same group. Vehicles' passage/presence/counting/occupancy/classification, and inference of microscopic and macroscopic traffic parameters including space and time headway, speed, traffic volume, and density are vital information that the fixed traffic sensors are supposed to supply. Nonetheless, the applications of VIP data are examined in a separate section due to the appearance of advanced vision techniques that have brought new achievements in traffic-related-object detection and tracking systems.

### 3.1. Operational mechanisms of traffic flow sensors

*1. Inductive loop detectors:* It is the most widely used traffic sensor, which is embedded in pavement by cutting a slot and placing wires in it. The detector contains a lead-in wire running from the main embedded wire loop to a power supply, and a lead-in cable that makes a connection between the detector system and a controller unit. The loop creates an electromagnetic field that is disturbed when a metal object such as a vehicle enters the field. The loop detector directly measures the presence and occupancy of passing vehicles while other traffic parameters including density, speed, and headway in both microscopic and macroscopic levels are deduced from the collected raw data. Size, shape, configuration, and settings of a loop-detector system vary upon the type of application and the speed limit of the roadway approaches. Further details on fundamentals of loop detectors are provided (18).

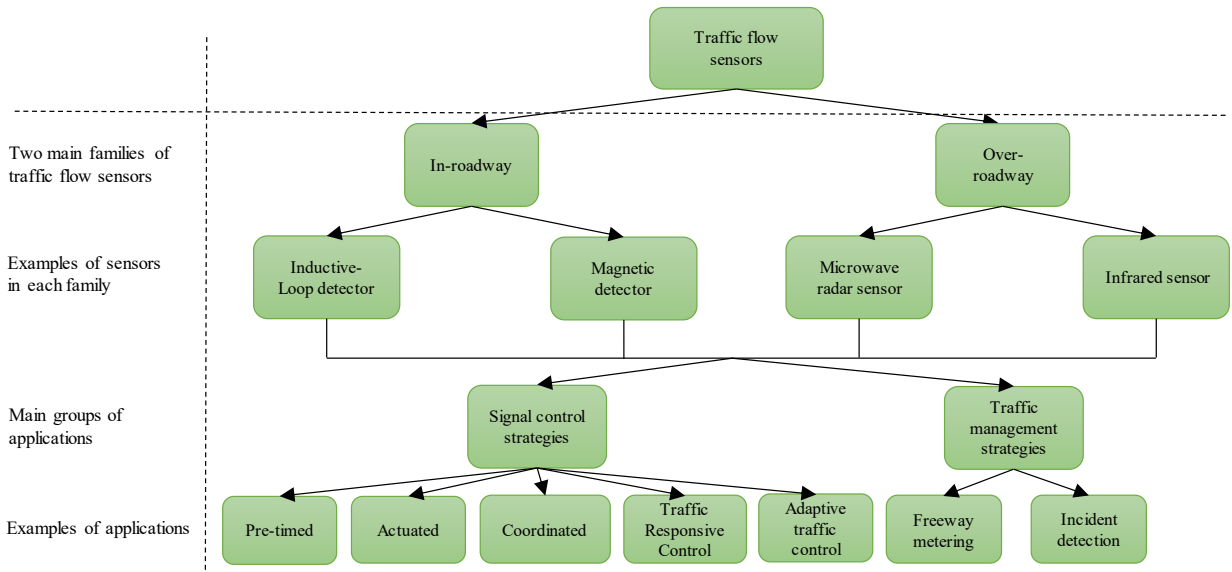

Fig. 2. Classification of traffic flow sensors and their associated applications.

2. *Magnetic detectors:* Magnetometer and magnetic sensors are two forms of magnetic detectors. These passive devices are operated based on perturbation induced in the normal Earth's magnetic field when a metal object such as a vehicle enters the magnetometer's detection zone. Such a disturbance is detected and counted once one vehicle is passing. Unlike the magnetometer that can detect both passage and presence, the magnetic sensors are unable to detect a stopped vehicle as changes in the vehicle's characteristics need to be observed for recognizing the vehicle presence.

3. *Microwave radar/Infrared sensor:* These sensors are operated by transmitting electromagnetic signals and receiving echoes from vehicles within the sensors' bandwidth. The major difference among these over-roadway sensors relies on the wavelength of the transmitted energy. While the microwave radar sensor utilizes the microwave wavelength, infrared sensors transmit energy in the near infrared spectrum. Infrared sensors are subdivided into passive and active types. The passive infrared sensors use the energy emitted or reflected from vehicles and roadways while the active ones, also called laser radar sensors, generate energy through laser diodes. When a vehicle is passing from the system antenna or laser diodes, a portion of the transmitted energy is sent back to the system receiver, which then the detection is made. Analyzing the frequencies of transmitted and reflected energy results in obtaining traffic flow parameters. According to the waveform transmitted by the sensor, the sensors are able to determine the passage and presence of a vehicle and calculate volume, lane occupancy, speed, vehicle length, queue, and vehicle classification.

### 3.2. Traffic flow sensor applications

The substantial increase in the traffic flow in transportation networks leads to the traffic congestion phenomenon, which calls for developing traffic control and management strategies in order to mitigate both recurring and non-recurring congestions. The recurring congestion refers to predictable conditions when demand exceeds the system's capacity such as daily rush hours; however, the nonrecurring congestion is usually caused by unpredictable incidents such as crashes, disabled vehicles, roadway debris, and weather-related issues. Traffic management and signal control concepts are the two major families of techniques for mitigating congestion effects. They essentially rely on appropriate traffic surveillance and detection systems for sensing and collecting traffic volumes, vehicular speed, travel time, and other features of traffic conditions. Ramp/mainline/freeway metering, lane management, incident detection, and traveler information systems represent demand or capacity management strategies for arterials and freeways. With regard to the traffic signal control methods, a variety of techniques has been developed for both isolated and coordinated intersections in a transportation network, ranging from pre-timed and actuated signal controls to advanced strategies such as the adaptive traffic signal control. In the following sections, the traffic management and control strategies are reviewed with knowing for the fact that a large portion of the necessary data for such strategies are collected through the above-mentioned flow sensors.

### 3.2.1 Traffic signal control strategies

Considering traffic characteristics and roadway geometry, an appropriate traffic signal control should be selected for defining the optimal cycle length, green time, and phase sequence as the major signal timing settings. Traffic signal controls for an isolated intersection usually involve a pre-timed or actuated mode or the combination of the two. As opposed to the pre-timed mode that major signal timing parameters are constant from cycle to cycle, the actuated control dynamically reacts to fluctuations in traffic demand in accordance with demand information collected by detectors, which results in having variable parameters in each cycle. The actuated operation is divided in three categories: semi-actuated, fully actuated, and volume-density control. In the semi-actuated control, detectors are installed only on the minor movements. Green time is always assigned to major approaches, except when a call is placed in one of minor approaches. This type is well-suited for highway operations with light traffic demand in cross streets and low-speed major approaches (19). Unlike the semi-actuated control, all traffic movements are actuated and equipped with detection systems in the fully-actuated control. This form of control is highly recommended for isolated intersections with high-speed approaches and variable traffic patterns during a day. The important features of each actuated phase contain minimum/maximum green time and passage time. The latter one indicates the fixed amount of time added to the green time of the phase on which a vehicle call has already been placed. This parameter can extend the green time of a phase up to the maximum green time. Also, the passage time specifies the maximum allowable gap between vehicle calls on a phase for retaining the green time. Volume-density control is essentially the same as fully-actuated with additional functions such as gap reduction and variable initial that allows the controller to have a variable passage time and a variable minimum green. In the gap production procedure, the passage time is reduced over time to a minimum value, which makes keeping the green time on a phase difficult as the phase gets longer. Variable initial chooses an appropriate minimum green time to serve the queued vehicles remaining between the stop line and the nearest upstream detector. Such a control is applicable in high-speed intersections.

Movements in multiple closely spaced intersections can be synchronized through the coordination concept, which is a strategy to provide a progressive traffic flow along an arterial in order to decrease the number of stops, delay, and travel time. The main objective in the coordinated system is to establish a time relationship between the beginning of the green time in adjacent intersections so that the amount of green time used by a continuous moving platoon of vehicles through the coordinated intersections is being maximized. Offset, the time difference between the downstream and upstream initiation green times, is the parameter that makes this time relationship between the coordinated phases. Coordinated traffic signals can be operated with both pre-timed and actuated modes. Intersections in close proximity with constant and large traffic volumes such as central business district and interchanges are well-fitted by pre-timed coordinated systems. Fusing semi-actuated and coordinated concepts generates a coordinated-actuated control which outperforms other signal operations in suburban arterials with heavy traffic (20). Additional information on the principal concepts, designing signal timing plans, and application of aforementioned traffic signal operation modes can be found in Traffic Signal Timing Manual and Traffic Engineering textbook (18, 21).

A group of intersections can be coordinated through a range of simple to advanced strategies. The key difference between them is how fast the controller adjusts the signal timing plan upon the traffic demand variation (17). The simplest way is to provide several timing plans for various times of a day or days of a week. For example, the timing settings of all signals in the coordinated group simultaneously change in three different times of a day: morning peak hour, afternoon/evening peak hour, and off-peak time. These timing plans need to be generated offline in advance based on historical data. This method is appropriate for locations with predictable traffic conditions when the same traffic patterns occur at the same time of day. However, occurrence of an incident such as inclement weather or a special event engenders unexpected behavior in the normal traffic conditions, which calls for a new timing plan beyond those pre-stored timing plans. To address such a challenge, traffic responsive and traffic adaptive systems have been developed to select the optimal timing plan that promptly responds to the real-time traffic demand.

In Traffic Responsive Plan Selection (TRPS), many pre-determined timing plans have already been input into a field master or a central computer system. Volume and occupancy data are continuously monitored by multiple detection systems that should be configured and set up to obtain an adequate representation of traffic conditions. The raw data are processed to calculate a set of parameters as a function of cycle lengths, splits, and offsets. These key parameters are compared with pre-defined thresholds to recall the best timing plan, from a set of pre-determined plans kept in a repository, which suits well with the current traffic condition. A fine-tuned TRPS selection can result in a smooth movement particularly in situations with significant unpredictable changes in traffic flows. Nevertheless, TRPS may come across with several issues such as inefficiency due to rapid offset transitions or late response to a quick change in traffic patterns since it takes time to calculate the parameters and set up a new plan. Adaptive Traffic Control System (ATCS) has been proposed to swiftly respond to traffic variations by predicting traffic conditions on a real-time basis; and then adapting signal timing parameters accordingly. A well-configured detection system is required to detect traffic volume continuously. Such a dynamic signal timing optimization, which results in the utilization of the maximum capacity, is very applicable when a high traffic volume fluctuation exists due to land use variations, an occurrence of incidents, or emergency cases such as preemption to ambulances. Examples of ATCSs are Real Time Hierarchical Optimized Distributed Effective System (RHODES), Optimized Policies for Adaptive Control (OPAC), Split Cycle Offset Optimization Technique (SCOOT), and Sydney Coordinated Adaptive Traffic System (SCATS). More details on key characteristics of the strategies as mentioned earlier for operating a group of intersections are available in Traffic Signal Timing Manual (18), chapter 9, and Traffic Detector Handbook (17), chapter 3. Excellent examples of algorithms for establishing TRPS and ATCS can be found in (22) and (23), respectively.

In addition to the cited applications, traffic signal control concept has also been utilized for special functions such as traffic signal preemption to the most important transport modes, traffic signal priority for transit vehicles and specified movements, as well as pedestrian signal actuation (18).

### 3.2.2 Traffic management strategies

Many offline and real-time strategies in traffic management systems are in need of traffic flow information collected through the detection systems. Examples of services that are supported by traffic flow sensors are incident detection, metering, congestion pricing, establishing traffic flow database for planning and evaluation purposes, and traveler information systems. It is worth mentioning that other traffic data sources can be involved in developing the mentioned services for escalating the accuracy and functionality. For the sake of focus on applications of the traffic flow sensors, we elaborate those types of management strategies that are contingent more on the traffic flow sensors' technologies. An early review on all kinds of management strategies can be found in Freeway Management and Operations Handbook (24).

1. *Freeway metering:* The crucial goal in freeway metering is to control traffic demand in such a way that the desired level of service and normal traffic conditions are obtained on the downstream of the control point, in which the detection system has been set up. The desired level of service gives rise to achieve flow maximization, distribution of total delay over the network, and increase in safety. Onramp, freeway-to-freeway, and mainline metering systems constitute the important forms of freeway metering. In these strategies, a signal control system with its detection systems is installed in the control point (i.e., on the ramp for the ramp metering, on the freeway mainline for the mainline metering, and on the connector ramp for the freeway-to-freeway metering) to manage the traffic demand (17). Onramp metering, for example, restricts the entrance of vehicles from an onramp to the freeway's mainline so that the maximum mainline flow rate is achieved without having a jam density in the downstream of the freeway. As shown in Fig. 3, a signal with a detection system can be installed on the ramp to disperse platoons of vehicles that are moving towards the freeway. The light is turned to green when the demand sensor identifies the arrival of a vehicle. After the vehicle is passed through the passage sensor, located after the stop line, the light returns to red for the next vehicle. Using a queue sensor at locations when ramp back up reaches the frontage road or surface street can handle the metering operation by either precipitating the vehicle passage or hampering vehicles' entrance to the ramp. Ramp metering can be implemented using three strategies: 1) pre-timed mode operated with a constant cycle length computed based on historical data, 2) local traffic responsive based on real-time traffic flow information, 3) coordinated traffic responsive control that analyzes the demand and capacity of the entire freeway to concurrently meter the flow rates of several ramps.

2. *Incident detection*: An incident is broadly defined as any non-recurring event that disrupts the smooth flow of traffic, generates an abnormal increase in demand, or reduces roadway capacity. Debris on the road, traffic crashes, highway maintenance projects, and special events are precise examples of incidents. Incident detection, as a critical stage in incident management, is the process by which the responsible agencies recognize an unexpected behavior in traffic flow characteristics through a variety of methods. The traffic measuring sensors are one viable option for detecting traffic anomalies (25). By applying the mathematical algorithms on the collected data, unexpected traffic features such as the propagation of a shock wave, variations in the average speed, density and occupancy in the both downstream and upstream of the suspected location can be discovered and labeled as the occurrence of a traffic incident. Traffic flows are streaming in a normal way far enough downstream and upstream of the incident location. As the flow perturbation is caused by both recurring and non-recurring reasons, the detection algorithms should be robust in distinguishing between the two. Next, an appropriate strategy is implemented within a coordinated program by various partners (e.g., State Police and State Department of Transportation) to promptly restore the smooth and safe traffic flow.

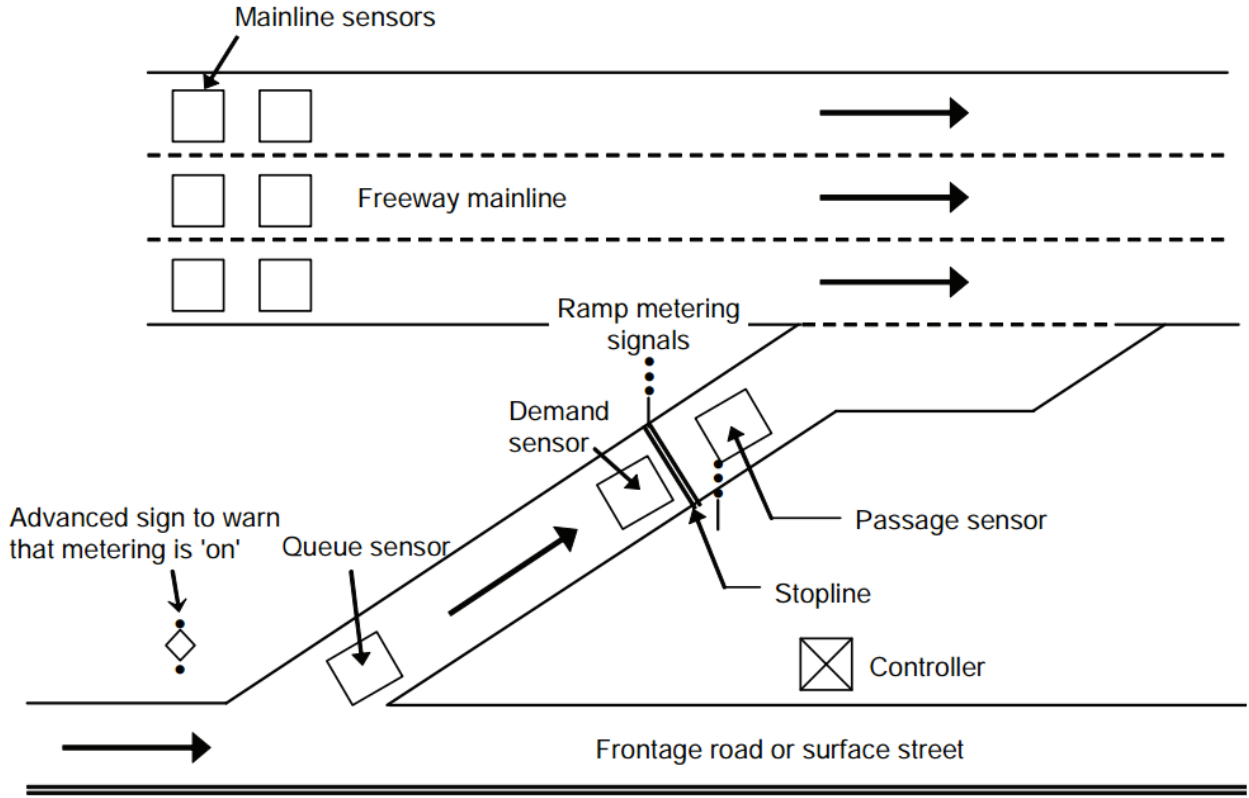

Fig. 3. Schematic of ramp metering detection system configuration.
(Source: Traffic Detector Handbook: -Volume II. No. FHWA-HRT-06-139. 2006)

### 3.3. Potential future research directions

Microscopic simulation models are indispensable tools to evaluate and replicate the stochastic nature of traffic conditions by considering the interactions between individual vehicles. These interactions are perceived by means of behavioral models such as car-following models, gap-acceptance, speed adaption, lane-changing, ramp metering, and overtakes (26). A car following model, as the most important and attractive behavioral model in the literature, describes the longitudinal behaviors of each vehicle with respect to its leading vehicle in the same lane. Much classical research in car following models have been conducted based on vehicle trajectory datasets collected through traffic flow sensors. With the appearance of car-following models, several commercial traffic microsimulation packages have been developed. VISSIM, for instance, utilizes a psycho-physical perception car following for modeling the drivers' traffic behavior. However, since the classical car following models have been introduced based on traffic flow theories in a closed-form mathematical functions, they have encountered troubles such as manual errors and inflexibility to involving additional information. One way to address these issues is to simulate the car following behavior directly from massive field data using data-driven techniques (27). Although the efficacy of a few data mining techniques on generating car-following models have been investigated, the performance of deep learning architectures in microscopic simulation models has not been examined yet. Furthermore, the current commercial traffic microsimulation packages have been constructed based on classical behavioral models rather than data-driven ones (28). Accordingly, a natural progression of microsimulation packages is to involve advanced data-driven based behavioral models in analyzing and simulating traffic patterns.

Alongside with the microscopic simulation packages, a wide range of macroscopic and primarily deterministic signal optimization tools (e.g., TRANSIT-7F and VISTRO) have been developed for optimizing signal timing settings of the mentioned control strategies. However, such programs are incapable of considering the stochastic nature of traffic conditions and uncertainties in the real-world traffic network including drivers and pedestrians' behavior, vehicles' characteristics, and variations in the traffic demand. In order to address this shortcoming, several researchers have integrated the optimization techniques with a microscopic simulation model, as the evaluation platform (29). Even though fusing an optimization technique with a microscopic simulator leads to having a more accurate model for generating signal timing plans, such a fusion procedure has been limited to offline applications due to the heavy computational burden. A future research direction is to investigate the feasibility of real-time applications of simulation-based optimization techniques for optimizing signal timing parameters.

Data collected using fixed-point sensors only present traffic states for the location where they were installed. Many ITS applications require the traffic information in a wider area. One cost-effective way to expand the spatial coverage of traffic information is to deploy other collective capacities such as participatory sensing (30). People and drivers, as moving sensors, can be involved in producing the spatiotemporal traffic-related data by leveraging the existing infrastructures (e.g., GPS, cellular network, and social media services) that have already been designed for purposes other than the traffic management strategies. The information from these technologies is aggregated through users' mobile phones and transferred to a central server by means of a global wireless network. On the other hand, with the emergence of Big Data and advanced data mining techniques, these heterogeneous data sources can be concurrently fused and analyzed with traffic flow sensors. This ameliorates the existing models that have been built up using a single fixed-point source. For instance, in addition to the incident detection models created based on traffic flow sensors, several algorithms for anomalous traffic patterns detection have also been proposed upon only one of other data sources, including geo-tagged social media data (31) and GPS-based trajectory data (32). Considering the participatory sensing and the advent of Big Data, all diverse data sources can be integrated into a single data fusion framework in order to create a more general and accurate model while overcoming the limitations of each source.

## 4. Video image processor

A video image processor (VIP) system typically consists of one or more cameras for taking vehicles' images and videos, a microprocessor for storing and processing images, and a software to apply computer vision and image processing algorithms for analyzing the raw image data and exploiting traffic parameters used in the traffic management (17). Traditionally, stationary VIPs have been categorized as over-road sensors since cameras are usually mounted on tall poles adjacent to a roadway or traffic signal mast arms. Mounting fixed cameras in a high position gives room to observing traffic movements across several lanes and multiple areas in one lane. Applying vision-based tools on images and videos from stationary cameras results in detecting and tracking vehicles, vehicle classification, vehicle volume and speed, counting turning movements, link travel time, etc. In-vehicle vision sensors, as the second type of VIP systems, can recognize the dynamic road scene while a vehicle is moving. Such systems, which are typically employed for in-vehicle safety applications and autonomous vehicle guidance, detect the actual state of the vehicle and its adjacent surroundings on a real time basis. Vision-based algorithms utilized for analyzing video images in stationary and moving cameras are almost disparate from each other (33). Tracking with 2D images and actual 3-D space in stationary and moving cameras, respectively, proves the major difference between the two groups of VIP sensors.

Notwithstanding abundant advantages coming from VIP systems such as their fast response, easy installation and maintenance, flexibility in changing settings, adding new detection zones, and providing rich traffic data, they suffer from a few weaknesses. Examples of shortcomings range from variability in the shape of objects and the high demanding computational time to being affected by weather conditions, illumination changes and occlusion (34). However, several studies have introduced novel techniques to address VIP issues in the detection and tracking process of traffic-related objects, which are described as follows. Afterwards, as illustrated in Fig. 4, the state-of-the-art of VIP traffic applications are categorized and manifested.

## 4.1. Vision-based algorithms

The major tasks of VIP systems are well-detection and well-tracking of traffic objects including vehicles, bicycles, pedestrians, traffic signs, roadway environment, and so forth (33). A majority of traffic parameters, and in turn traffic applications, are computed if vehicles are detected and traced with a minimum error. Detecting the presence of vehicles in designed detection zones leads to obtaining traffic volumes by recording vehicles' passage in each lane. Locating a vehicle imparts information on the travel time between two locations, average link speed, and counting turning movements.

With regard to the object detection, vision-based algorithms require meeting a variety of challenges in detecting traffic-related objects (34, 35). Vehicles, as an important traffic object, are substantially different in size, shape, and color. Also, environmental changes such as day-time, illumination conditions, existence of unpredictable objects, and occlusion due to shadowing affect the detection process. Operating on a real-time basis, particularly for driver-assistant applications, is another demanding challenge that requires to be addressed for building a reliable and accurate object detection systems.

For attaining a rigorous and exact detection model, many subtasks should be taken such as detecting a region of interest, background subtraction, removing an occlusion, object's features detection and description, as well as classifications. Studies in the traffic-object detection process can be categorized into three groups. The first group consists of seminal studies in the computer vision field that have postulated the fundamental techniques for processing images (e.g., smoothing images and edge detection, feature detection and description such as SIFT and SURF, machine-learning based classification such as Adaboost cascade of simple features, and deep convolutional networks for image recognition). A large volume of published studies has sought to combine and integrate the fundamental techniques with a specific minor modification in algorithms to come up with a more robust framework for the object detection. In the last group, several studies have applied the proposed object detection structures in transportation systems to obtain specific goals like detecting vehicles and pedestrians.

Several survey studies have presented different techniques for classifying the objects on the roadway, where the comprehensive reviews are available in the references (33-35). A high-level view of the object detection process is to distinguish between the subject object and its environment. The algorithms strive for, first, identifying the potential locations that the object may exist and secondly verifying the hypothesis in the first step (34). Both steps are implemented based on motion and appearance features of the objects. Appearance features originate from prior knowledge about the particular object and include symmetry, geometrical edges such as corners and horizontal/vertical edges, Histogram of Oriented Gradient (HOG), and Haar-like features, shadow, vehicle texture patterns, and vehicle lights (34, 35). Unlike the appearance-based approaches, motion-based approaches obtain relative motion of interesting points using optical flow calculations. The whole idea of the optical flow is to determine the pixel intensity change over time. When an object comes into/out of the background, a noticeable intensity change occurs which can be measured using the optical flow theory.

The objects' distinctive invariant characteristics are extracted and learned using interest point detection and description techniques such as Scale-Invariant Feature Transform (SIFT) (36) and Speeded-Up Robust Features (SURF) (37) from a training set of images. Then, the objects are recognized by training a machine-learning classifier such as boosting algorithms, nearest-neighborhood algorithms, neural networks, and SVM that matches the image feature descriptor to a database of features from known objects. As a separate or complementary step to the matching step, the clustering and image segmentation analyses are performed to identify the clusters belonging to the target object using the probability distribution of the object.

Nonetheless, the hand-crafted features (e.g., HOG, Haar-like, SIFT, and SURF) might be vulnerable to traffic and environmental conditions. One way to overcome these issues is by utilizing the deep learning schemes that are capable of automatically driving high-level features and learning thousands of objects from a pool of images. Convolutional Neural Network (CNN) is a successful example of deep learning architectures that is able to make almost correct assumptions on the nature of images (38). Although the key idea in the CNN is similar to the ordinary feed-forward artificial neural network, they differ in sense of connectivity patterns between the neurons in adjacent layers. The CNN takes the advantage of the spatially local correlation by connecting neurons to only a small region of the preceding layer. Such a local connectivity between nodes leads to having the smaller number of weights, which mitigates the curse of dimensionality and overfitting problem. Notwithstanding the existence of various CNN extensions in the literature (39, 40), a typical CNN architecture constitutes a sequence of layers (e.g., input layer, convolutional layer, pooling layer, fully-connected layer, and dropout layer), in which each layer transforms an input volume to an output volume of neurons using a set of operations. Finally, the extracted high-level features from the sequence of layers are fed into the last fully-connected layer for performing the classification task.

Vehicle detection is generally followed by a tracking process to not only improve the detection process but also identify the vehicle's motion characteristics. Because vehicles have temporal continuity in sequential frames, the vehicle location can be predicted using the past movements, which leads to augmenting the accuracy of the hypothesis generation in the aforementioned detection process (34). Dynamic tracking parameters of the objects, including speed and acceleration/deceleration in terms of pixels or meters, are measured by applying data association across frames and features that have already been extracted for the detection process (35). Template matching, optical flow, Kalman filtering, and particle filtering are the common methods for the object tracking. In (41), as an example of the template matching, after detecting a vehicle, a template region of each vehicle is extracted by finding the distance between the main features and the vehicle center. The template area is compared with the template matching for tracking. Feature-based tracking is another approach where the position estimation can be smoothed using filtering techniques (42). For instance, the object tracker utilizes a set of Haar and Triangle features for stabilizing detection and tracking. A weighted correlation is carried out to find the features that have a high impact on the matching process while Kalman filtering is integrated to refine the detection process and decrease false positive rates. In (43), the particle filtering is used to track vehicles by predicting their states and updating observation. In the meantime, integrated likelihoods such as estimated probability of vehicles are implemented to maintain the tracking accuracy and robustness.

## 4.2. Applications

### 4.2.1 Moving objects' detection, tracking, and classification

Traffic-moving objects consist of vehicles, bicycles, and pedestrians. According to Federal Highway Administration, the term vehicle is generally understood to mean motorcycles, passenger cars, buses, and trucks with a different number of axels. As cited above, detection and tracking are the two vital steps towards many video-based traffic applications. Both speed and red light cameras, which are able to detect vehicles and their license plate, are used at intersections and highways to detect drivers who exceed the speed limit or enter the intersection after the light is turned red. Installing these cameras under the law enforcement programs improves driver compliance to traffic regulation and enhances safety by reducing accident risks.

Roadway traffic monitoring and autonomous vehicle guidance are two major applications operated based on vehicles' detection and tracking as well as their environmental components such as road geometry, lane, and center line (33). Processing video images collected through stationary cameras leads to monitoring traffic stream on a real-time basis by measuring and updating instant and average traffic flow, turning movements at intersections, speed distribution measurements, queue lengths, and vehicle counts for various vehicle's classes. Transmission and further analysis of such rich traffic state information in Traffic Management Center also bring benefits on optimizing traffic signals, updating traveler information systems such as variable message signs, evacuation management, and minimizing the travel time for emergency vehicles.

On the other hand, in the automatic vehicle guidance, moving cameras mounted in vehicles are employed for determining the vehicle relative position to the lane, location of other obstacles and moving objects with respect to the subject vehicle, and broadly capturing the vehicle and its road environment (33). Thus far, several studies have tested the efficacy of computer vision algorithms to provide perception systems for autonomous vehicles. In pursuit of moving towards fully autonomous vehicles, Advanced Driver Assistant Systems (ADAS) have been designed to raise users' safety and comfort yet decrease economic costs and pollution (44). Various types of sensors may deploy in ADAS to collect necessary information for informing drivers about their surroundings and performing repetitive tasks. Cameras, GPS, ultrasonic sensors, radar sensors, and LiDAR are the examples of sensors in ADAS. Vision-Based DAS (VB-DAS) constitutes the core technology in self-driving vehicles for analyzing the traffic scene around vehicle, which consists of multiple cameras and a processing unit (45). After receiving the raw video images, VB-DAS primarily relies on advanced computer vision algorithms to dynamically analyze the vehicles' surroundings. Examples of ADAS are adaptive cruise control (46), automatic parking assistant system (47), emergency lane departure and collision avoidance system (48), night vision systems (49), and so forth. Interested readers on VB-DAS are referred to the several survey articles that have been examined various aspects of VB-DAS and autonomous vision. For example, Klette et al. (45) provided a high-level review on safety and comfort functionalities of recorded video data (e.g., avoidance of blind spots and night vision), basic perception of traffic environment (e.g., distance computation, motion computation, and object detection and tracking), and VB-DAS examples. In-depth and comprehensive review of the state-of-the-art on various tasks in autonomous vision (e.g., recognition, reconstruction, motion estimation, tracking, scene understanding, and end-to-end learning) is provided by Janai et al (50).

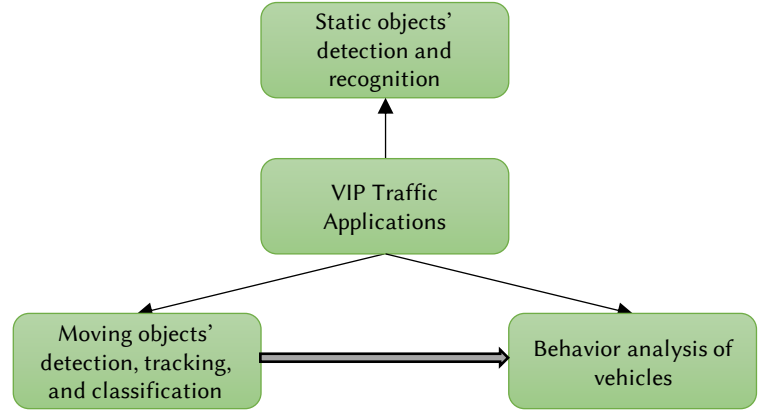

Fig. 4. Main applications of VIP in transportation systems. The gray arrow demonstrates the interdependency between applications.

Pedestrian protection is another particular task of ADAS accomplished by using a reliable on-board camera and a pedestrian detection system. Protection systems are capable of detecting stationary and moving people in the interest region to warn the driver for an appropriate action to prevent the car-to-pedestrian accident (51). Analyzing pedestrians' dynamics and behavior assists the prediction of possible accidents between vehicles and pedestrians (52).

In addition to pedestrians, cyclists are another group of transportation system users who are classified as the most vulnerable road users since they are at risk of suffering injuries in the case of accidents with larger vehicles such as private cars. By using the in-vehicle vision systems, detecting and tracking bicycle movements, and in turn keeping bike riders' safety, have become easier (53). As bicycles share the transportation network with other transportation modes (54, 55), their traffic characteristics need to be taken into account. Counting bicycles is necessary information for traffic managers to assign sufficient lanes for riding bikes (56).

### 4.2.2 Static objects' detection and recognition

Traffic signs, installed at the side or above roadways to convey critical safety-related information to road users, are one of the static traffic-related objects. The Manual for Uniform Traffic Control Devices (MUTCD) requires transportation agencies to adopt assessment strategies to maintain the sign visibility at standard levels (57). Traffic sign recognition, as a subsystem of ADAS, plays an important role in increasing traffic road safety by informing drivers regarding traffic rules, navigating drivers to the destination, and alerting drivers to potential hazards on the network environment. Traffic sign recognition makes the procedure of sign maintenance more efficient without needing to manually perform the sign-condition evaluation (58). Designing traffic signs with standard colors and shapes increase their chance of being recognized through the vision-based algorithms, although the unpredictable traffic environment and traffic sign deformation are obstacles to developing a reliable system. Surveys on the traffic sign recognition systems and their details are available in the references (58, 59).

Many applications in ITS have been implemented using accurate License Plate Recognition (LPR) systems including toll payment, parking fee payment, traffic law enforcement, link travel estimation, and road traffic monitoring (60, 61). Furthermore, an LPR system has been used in developing the vehicle manufacture and model recognition scheme, introduced as a novel detection and classification system compared to those that only classify objects to vehicle or non-vehicle (62). In the LPR systems, the integration of computer vision and feature extraction algorithms is being applied on still images or some frames of the image sequence to obtain temporal information that is effective in improving recognition process. In the LPR systems, a real-life application needs to meet a broad spectrum of challenges such as a quick processing time under various environmental conditions and analyzing plates from different color and languages (60). A comprehensive review on existing LPR techniques explaining the pros and cons of each method is available in the reference (60).

### 4.2.3 Behavior analysis of vehicles

A robust vehicle detection and tracking system deliver rich information on spatiotemporal features that are used to learn, model, and predict the behavior of vehicles on the road. According to (35), the vehicle behavior characteristics are categorized based on the context, identification of pre-specified maneuvers, as well as vehicles' trajectories and classification. Considering the role of context, characterizing the on-road vehicle behavior leads to classifying the driving environment into intersection and non-intersection roads, predicting the long-term behavior of vehicles in roundabouts, and analyzing the turning movements at intersections. Detecting behaviors such as overtaking and undertaking, normal and abnormal, exiting and merging, car following and lane changing are exemplified as the maneuver identification category. Vehicle trajectories, a long sequence of vehicle velocity and position over a time period, are used to predict the long-term vehicle movement in the urban and highway roads, signalized intersections, and roundabouts.

Learning the detailed vehicular/pedestrian movements, which can be achieved through clustering homogenous trajectories, is valuable information that uncovers the characteristics of interactions between traffic objects, and vehicle/pedestrian travel patterns and behaviors. Such rich information is employed in many applications such as motion planning systems for autonomous vehicles (63), predicting pedestrian counts for emergency response systems (64), traffic accident estimation through analyzing car-following and lane-changing behavior, and calculation of time-to-collision (65).

### 4.3. Potential future research directions

The future trend in VIP-based traffic applications has primarily been focusing on improving the computer vision technology so as to develop more robust and accurate object detection and counting systems. A robust and reliable object detector is able to detect all pre-defined object classes regardless of environmental conditions, illumination changes, pose changes, image scaling, translation, and rotation. For real-time applications, the detection scheme should also be fast enough with a low computation cost while not sacrificing other superior features. However, many proposed object detection systems are not able to meet all the requirements. For example, increasing the computational speed of a real-time vehicle detection system may give rise to decreasing the number of the vehicle classes that the system can distinguish. Accordingly, the current and future research endeavor for traffic-object-related detection systems in ITS applications must foster and provide the aforementioned objectives so as to not only detect various components of a traffic scene (e.g., lane marking, road furniture, different types of current and future vehicles and buildings) in a fast and efficient way, but work well under various scenarios (e.g., variable resolution, inner-city or highway traffic, and rain in night) (45).

Autonomous vehicles are an ongoing ITS research field that is in dire need of vision-based detection systems for sensing its environment. Although a vast majority of research in ADAS has made a great success in the field of autonomous vehicles, generating fully autonomous vehicles is still decades away. The major obstacle in developing fully autonomous vehicles is the lack of very accurate and robust perception systems to fully recognize complex dynamic environments (50). Despite many advances in object detection and tracking systems, most of the current VB-DAS are still incapable of distinguishing between various objects in an error rate level that is acceptable for permitting fully autonomous cars. This calls for more effort to manufacture an accurate and robust control system in autonomous vehicles, where the VB-DAS is a major component of such a system, that is capable of operating well in many possible scenarios.

One active research in computer vision for overcoming the current deficiencies in VB-DAS is the use of deep learning architectures in VB-DAS tasks such as object detection and tracking. The promising goal of deep learning algorithms is to learn and extract image representation features automatically rather than using the hand-crafted features designed by a user. However, in spite of demonstrating its outstanding performance in the object detection systems (66, 67), it is not clear why these deep learning architectures outperform traditional methods that operate based on hand-crafted features. In other words, the black-box nature of deep learning architectures gives no insight into the structure of the function being approximated. As a consequence, in order to improve such models intelligently rather than with the trial-and-error approach, it is required to fully understand the nature and evolution of the learned representation in each layer (68).

## 5. Probe vehicle and people data collection

The fixed-point sensors enable to only measure temporal information in a particular location or a small area, which results in lacking spatial traffic information that may not be representative of the network as a whole. One way to address this issue is to use vehicles and individuals that are equipped with location and communication providers as "probe vehicle and people" to sense their spatial information when they are performing their regular trips. In this study, we examine three well-established positioning tools that record spatiotemporal information of floating sensors while moving in the network. These technologies are listed as Global Positioning Systems (GPS), mobile phone cellular networks, and Bluetooth. Devices equipped with these technologies (e.g., drivers' smartphones) can track vehicles or people's locations indexed in time to create their trajectory data. Such spatiotemporal data are worthy inputs for a variety of transport-domain applications that are displayed in Fig. 5 as an overall view and will be set out in the following sections. Before elaborating the related applications, we briefly characterize the mechanism of each technology.

### 5.1. Global Positioning System

#### 5.1.1 Operational mechanism of Global Positioning System

GPS, also called NAVSTAR by the US Department of Defense, is a satellite-based navigation system built up by 24 satellites that are orbiting the earth. Satellites are constantly moving and powered by solar energy while backup batteries supply their energy when there is no solar power. However, each satellite life is roughly ten years which necessities a constant replacement process. A GPS contains three parts: the satellites in the earth's orbit, the ground control stations, and the GPS-receiver devices such as mobile phones or handled GPS units (69). At each location on earth, at least four satellites are visible. A GPS device receives the radio signals broadcasted by the satellites. Radio signals transmit information about the current time and position of the satellite. Knowing the fact that the signal's speed is the same as the light's speed, the GPS receiver compares the difference between the transmitted time and receiving time to calculate how far a satellite is from the user's location.

Calculating the 2-D position (i.e., latitude and longitude) requires obtaining the signals from at least three satellites. The GPS device can pinpoint the user location by calculating the distance from these three satellites through a process called trilateration. A user is located somewhere on the sphere surface with a radius equal to the distance from the first satellite. If the user knows the distance from the second satellite, the overlap between the first and second sphere (i.e., a perfect circle) determines the possible locations of the user. Finding out the distance from the third satellite results in two locations at intersections between the perfect circle and the third satellite. Since one of the two points is located outside of the earth, another point is labeled as the user's location. Yet approximating the location using only three satellites' information may not be exact enough. More accurate and precise location and also the 3-D position (i.e., latitude, longitude, and altitude) require four or more satellite signals. Thus, the object's position coordinates and its corresponding timestamp over a period is able to be collected using a GPS-equipped device.

In spite of emerging advances in the GPS technology, various factors affect its accuracy in identifying objects' position. The inability of signals to pass through solid objects such as tall buildings is one of the major error sources. In a crowded area with huge skyscrapers, the signal is distorted from its main direction before reaching the receiver, which in turn results in increasing the travel time and generating errors. Orbital error, relative geometry of satellites with respect to each other, and the number of visible satellites are some of the other sources for errors.

*5.1.2 Transportation-related applications of GPS*

GPS is one of the main area-wide sensors for generating the massive movement/trajectory data of moving objects such as people and vehicles. Although it was first developed for US military services in the 1970s (69), developments in the GPS technology have motivated research institutes and authorities to use this system in many other applications such as traffic and transportation systems. A GPS trajectory, also called movement, of an object is constructed by connecting the GPS points of its GPS-enable device such as a smartphone. A GPS point, here, is denoted as (x, y, t), where x, y, and t are latitude, longitude, and timestamp, respectively. The original object trajectory is continuous while its corresponding data are discrete, which necessities linear or non-linear interpolation for finding positions between two consecutive discrete data points. Such a series of objects' chronologically ordered points can also be viewed as a set of spatial events. Spatial events are defined as consecutive occurrences of certain movements' features that are localized in the space (70); hence, its real definition may vary according to the specific interest of the researcher.

Vehicle fleet management and monitoring in urban and suburban areas can be considered as initial practical usage of the GPS technology (69). The installed GPS-enabled device in emergency, police, and transit vehicles monitors and transfers the vehicles' movement information, also well-known as Floating Car Data (FCD), to a control center via a wireless communication device. Such an operation can provide several advantages to the vehicle fleets such as detecting any traffic violation and improving the disposition of clients' orders in taxi companies. Several researchers have collated FCD to recognize current traffic states and predict its near-future conditions as the main components in designing and implementing advanced traveler traffic information systems and advanced traffic management systems. The traffic information calculated by processing and analyzing FCD consists of one or multiple layers including short-term link speed and travel time prediction (71), route travel time distribution (72), traffic incident detection (73), congestion detection by estimating traffic density (74), dynamic routing and navigation (75), route choice behavior (76), map-matching and path inference (77), and dynamic emission models (75).

Human mobility behaviors and their relation to traffic conditions can be characterized by constraining their movements into the underlying street network using the GPS data (78). Analyzing the people's mobility under a large street network leads to understand and predict the traffic distribution using a number of GPS recorded points in each street link. For example, the number of positions in an individual street link presents how congested the link is. Also, a power-law behavior can be fitted to the observed traffic using the speed effect of vehicles extracted from their GPS data (78). It is worth mentioning that the power-law distribution is one of the famous models that has been observed for the human mobility movement from real-world human mobility datasets in a broad spectrum of studies (79). Studying urban mobility benefits to characterize the attractiveness of various locations in the city, which is a significant factor in developing trip attraction models in the urban planning (80). Taxi GPS-based trajectories have utilized to infer driver's actions as a function of observed behaviors such as predicting actions at the next intersection, providing the fastest route to a given destination at a given departure time, and predicting the destination given a partial route. Moreover, vehicles and human's mobility studies based on positioning data have injected a variety of advantages in terms of inferring significant locations, modes of transport, trajectory mining, and location-based activities (81), which are discussed as follows.

Locations with the high frequent visits or high individual's dwell time is labeled as significant or interesting locations such as office buildings or intersections. As the object's movement is a spatial-temporal phenomenon with a sequence of events, significant locations are extracted from the occurrence of events. First, the events that happen repeatedly are detected and extracted, then relevant places to each group of events are determined using spatial-temporal clustering algorithms. Analysis of the aggregated events and places leads to determining the significant locations pertaining to the subject application (70). Considering the fact that GPS receivers almost unable to get signals in the closed environment, office buildings as an example of significant locations can be detected by clustering the GPS points (82). Furthermore, future movements of individuals can be found through efficient probabilistic algorithms such as Hidden Markov Model.

Inferring a user's motion mode can guide not only others to utilize an appropriate mode for a specific journey but themselves to understand their own life pattern. With regard to the application systems, discovering congested/uncongested traffic conditions and proper modes for various origin-destination empower traffic management systems to smooth urban mobility. Mining of GPS-based users' trajectory leads to deducing users' mode(s) of transport (e.g., bus, walk, car, bicycle, and train). Detecting the mode of transport primarily depends on the movement features that distinguish between modes. After dividing the GPS track into uniform segments, basic features of each segment including length, mean velocity, and maximum accelerations are employed to recognizing the transport mode(s) (83). Nevertheless, speed and acceleration are erratic and unpredictable due to changes in traffic conditions, weather, and types of roadway. The vulnerability of basic motion characteristics forced researchers to introduce additional features such as changing rate in direction, stop rate, and velocity change rate (84). Yet the hand-crafted features may still cause some major drawbacks including vulnerability to traffic and environmental conditions as well as possessing human's bias in creating efficient features. One way to overcome these issues is by utilizing the Convolutional Neural Network (CNN) schemes that are capable of automatically driving high-level features from raw GPS trajectories (85). Moreover, transportation network information including real-time bus locations, rail line and bus stop spatial information, as well as capacity and location of parking lots are helpful features for understanding the transition between modes since these are the locations that people often change their current mode (81, 86).

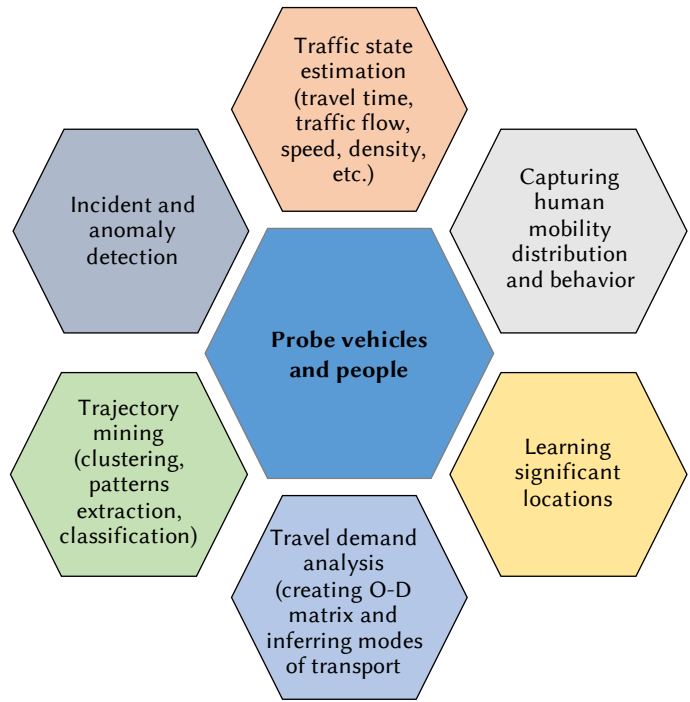

Fig. 5. Main applications of probe vehicle and people data in transportation systems.

Vehicle and human trajectory mining can be achieved by converting their movement into a sequence of spatial locations (i.e., a set of representative points). Trajectory clustering, trajectory patterns extraction, and trajectory classification comprise the three most important groups in the trajectory mining (81). In clustering, typical movement styles that are relatively close to each other are discovered and grouped into one cluster (87). The measure of closeness is considerably contingent on the research application; yet computing the distance between trajectories on the basis of temporal and spatial information is a standard technique to measure the closeness (87, 88). In the trajectory pattern extraction problems, frequent (sequential) patterns, which are the routes frequently followed by an object, are discovered for predicting the future movement using association rules while the trajectory is viewed as spatiotemporal items (89). Considering the frequent trajectory patterns and users' context as features, vehicles' paths can be classified using associative classifications (90). The trajectories' labels are composed of every type of mobility behavior in the road network including dwell time in the current place, destination (next place), trip purpose, or activity such as shopping, parking a car, dining, etc. In addition to associative classification methods, aforementioned behaviors can be learned from movement data using a conditional probabilistic framework (91). Useful transportation and traffic planning applications based on GPS trajectory mining are exemplified as a carpooling system, prediction of the next destination or the current place stay duration (91, 92), social networking service for sharing users' life experience, and travel recommendation systems (93, 94).

Traffic anomaly detection can be categorized as a study either in the traffic incident management or a type of trajectory mining. Outlier/anomaly is regard to any behavior that not conform to the expected and normal behavior. GPS-based traffic outliers are grouped into two groups. The first group is referred to detect a driver's trajectory or a small percentage of drives' trajectories that their behavior is considerably different from others such as a fraudulent taxi driver or a traffic offender. An outstanding example of this category is the work of (95), where a fraudulent taxi behavior, who overcharges passengers for more than the actual distanced travel by altering the taximeter into a lower scale, is discovered by comparing its real speed based on GPS logs and the speed obtained from a malfunctioned taximeter. Trajectory data of GPS-equipped transit vehicles have also been deployed as real-time information to automatically recognize the urban road network changes such as a construction process or newly built roads (32). The second group investigates an occurrence of traffic anomaly by drawing a distinguish between a vast majority of trajectories. For instance, traffic anomalies caused by accidents, disasters, and unusual events are captured according to irregular drivers' routing behaviors on a sub-graph of a road network. Comparison between current suspicious drivers' actions is carried out with remaining trajectories at the same time and location or with the historical record of drivers (96). Using travel information systems, the detected outlier in the system is reported to the traffic management system for performing appropriate actions; in the meantime, drivers are being informed on the occurrence of ahead traffic jams, which may result in re-routing.

## 5.2. Mobile phone cellular networks

### 5.2.1 Operational mechanism of cellular networks

There are a variety of ways to find the location of a mobile phone, specifically smartphones that are prevalent on the market. The most common and widely used technologies to trace cell phones are summarized as GPS, cellular networks, Wi-Fi, and Bluetooth. The goal in this section is to unravel how to retrieve the location of a mobile phone at a given moment using signals it sends to cellular networks.

A mobile phone network, also called a cellular network, is a wireless network constituted by a set of cells that covers a land area, where transmission of voice, text, and other data is carried out through at least one fixed-location Base Transceiver Station (BTS) allocated to each cell. Multiple BTS in close proximity are managed with one Base Station Controller (BSC). Thus, a huge number of cell phones can communicate together even when they are moving through more than one cell. Today, the cellular technology, which is communicating directly with a ground-based cellular tower, has been implemented in almost all mobile phones. A broad digital cellular standard has been introduced and used across the world. The most market shares for 2-G, 3-G, and 4-G digital cellular networks are Global System for Mobile Communications (GSM), Universal Mobile Telecommunications System (UMTS), and Long-Term Evolution (4-G), respectively. Here, the mechanism of the GSM cellular network, as an example and the highest market share standard, is briefly reviewed.

When a mobile phone is connected to the BTS by exchanging signals, the network operator approximates the mobile phone location according to the coordinates of BTS, which results in having an accuracy at the Cell-ID level. Although this is the most used technique for locating the mobile phone, other techniques such as Received Signal Strength method (RSS) and triangulation have been proposed in literature to increase the accuracy; however, these methods acquire extra network elements that are not essential for cellular networks to work.

In addition to the location information, parameters regarding the rate of using cellular networks are helpful in determining traffic parameters (97). Handover is the mechanism that provides a permanent connection by switching an on-going call between different BTS while a mobile phone is moving through multiple cells in the network. The Cell Dwell Time (CDT) shows the duration that a phone remains with a BTS before handing off to another base station, which is an index to compare the level of traffic congestion between cell areas. Another parameter for understanding crowdedness in a zone is the Erlang that indicates how many hours a user has utilized the network. One Erlang equals to one-person hour of phone use. It should be pointed out that all network data can be collected even when the phone is only switched on.

*5.2.2 Transportation-related applications of cellular networks*

Cell phones as a pervasive sensing device and enthusiastic adoption across most socioeconomic strata is a reliable source to record users' spatiotemporal information. Such capabilities on mobile phone data have persuaded urban researchers and authorities to employ this real-time and massive location data on a variety of behavioral applications, ranging from inferring friendship network structure and real estate market, to deriving a geography of human urban activity and emergency management (97). Besides, analyzing and modeling commuters' movement turns these large-scale spatiotemporal data into unprecedented and actionable insights in the field of urban planning and traffic management.

The main advantageous of mobile phone data, compared to other positioning data sources, is the ability to gain large data on the individual level and having an enormous market penetration rate in a large number of countries. According to Statista, the mobile phone penetration is forecasted to reach 67 percent of the population worldwide by 2019 when smartphone users have taken the half of mobile cellular subscriptions. The mobile phone is a ubiquitous computing device that nearly all time is relatively close to the user. Also, its high penetration rate makes the mobile phone data cover a metropolitan area, which as a consequence enable users and traffic management centers to understand traffic conditions over a wider area.

Mobile phone data, utilized in transport domains, can be divided into two parts: 1) spatiotemporal information which identifies the location of a phone over a period, 2) data related to the usage rate of the cellular network such as handover, Erlang, and the number of calls. One should pay heed to the fact that as long as location-based applications are the concern, it does not matter what technology has been deployed to obtain vehicles and commuters' flow and trajectory data. As mentioned, a mobile phone is located by means of various techniques. A huge growth in the market share of phones with built-in GPS receivers gives room to detect the location of a phone using GPS, which is a more accurate technology with higher resolution. Yet the cellular networks technology is still an effective way to capture real-time, cellular-signal data points; particularly, where the GPS technology is not capable of collecting movement data such as inside buildings or once the mobile phone's location service has been turned off by the user. Moreover, on the premise that at least one cell phone is carried inside a moving vehicle, individuals' and vehicles' movements data are collected more easily for cost-effective and innovative traffic studies. *Consequently, all the GPS-based travel patterns and transportation trends mentioned in the section 5.1.2 have also the potential to be inferred from the movement data provided by the cellular network even though the differences in the spatial coverage and resolution of the two systems should be taken into account*. Thus, we focus more on potential transport-related applications of cellular networks that collate network-usage rate parameters rather than location-based information.

Traffic parameters such as origin-estimation (OD) trip matrix, travel time, speed, traffic flow, traffic density, and congestion are obtained by applying a proper procedure on the mobile phone data (98). The final goal in the OD research area is to estimate or update the average number of trips going between each pair of traffic zones, called origin and destination, during a period of time. The location of on-board cell phones can be determined in terms of the cells they travel through it. The centroids of cells or the location of BTS in each cell is considered as the location of the mobile phone in each cell (99). Traffic volume is a useful parameter for understanding the traffic patterns and demand in roadways. Cellular systems can provide the traffic volume information by counting the number of vehicles moving in each cell area. Also, a transition between every two cell borders can be detected using cellular network parameters such handovers, since the handover works the same as a virtual traffic counter between borders of two cell areas (100). The most important advantage of cellular systems compared to other methods is no need for installing additional infrastructures such as loop detectors and VIPs. Traffic speed, usually indexed as the time-mean and space-mean speed, is a parameter that shows the road level of service. VIPs and loop detectors are the methods for measuring the speed, which incur the installation and maintenance costs as well. In cellular networks, the speed can be computed by measuring the location of the phone periodically. Then, having the distance and travel time between two consecutive location leads to measuring the speed (100). Double handovers is another way for measuring the speed. If a call is sufficient long to produce two handovers, which means passing through two cell areas, speed is estimated using time and location information that a vehicle enters and exits the boundaries of a specific cell. The similar handover concept in the speed estimation has been deployed for measuring the travel time. The time information on two adjacent handover occurrences identifies the travel time of the cell area that a phone crosses. However, as the cell area covers multiple road sections, a reliable mapping process requires matching the captured travel time to the real road section that the phone has passed.

Traffic incident management is another area that can be addressed using cellular systems (97). First of all, any sudden change in above-mentioned traffic parameters such as speed and travel time is a clue for an incident occurrence. Furthermore, comparing the current rate of mobile phones' usage in a specific location with the historical average rate usage in the same location is a sign of a congestion situation owing to the facts that not only the number of phone users is more in heavy traffic but users intend to communicate more with people at their trip destination. Such analysis can be fulfilled using the Erlang index. CDT, which shows the duration of being registered to a specific BTS before a handover occurs, is another measurement for identifying the level of traffic congestion. High CDT of a phone inside a probe vehicle indicates a long travel time and in turn a high level of congestion in the phone's cell area (101). Furthermore, as Erlang is a measurement of the mobile phone usage in a sector, it can be used to estimate the number of users in a cell area, which is highly correlated with traffic density (98). Comprehensive reviews on studies that have developed frameworks and methodologies to characterize transportation system parameters through cellular network data are available in the references (97, 98).

*5.3. Bluetooth*

*5.3.1 Operational mechanism of Bluetooth*

Bluetooth is a short-range, wireless technology that uses a short-wavelength radio communication system for exchanging data between Bluetooth-equipped devices such as mobile phones and car radios. Such a standard wire-replacement protocol simplifies connection settings between devices in terms of security, network address, and permission configuration. It follows a master-slave architecture, in which a master device is communicating with one or more devices (slaves) for transferring data over a short-range and ad-hoc network in the range of 1 to 100 meters, dependent on the device's Bluetooth power class.

The Bluetooth traffic data collection system leverages Bluetooth probe devices, also called Bluetooth detectors, located adjacent to roadways for scanning Bluetooth-equipped devices inside vehicles such as driver's smartphones or built-in Bluetooth car audio. The probe device detects and records the unique Machine Access Control (MAC) address of Bluetooth-enabled vehicle that has entered to the probe device's radio proximity. The MAC address is a 48-bit physical layer address which is served as an effective and unique identifier of each Bluetooth-enabled device. With knowing the location of the installed probe device, the presence location and time information of those Bluetooth-enabled vehicles that has just crossed the probe device are also identified. Using a sensing system with multiple probes in sequence, vehicles' trajectories can be built by recording the MAC address of their on-board Bluetooth-enabled devices at multiple stations and extracting the unique MAC address in a chronological order. The stored information is then transmitted to a control server for further analysis and uses. Type of probe vehicles, the required number of probe vehicles, and the location of probe devices in transport networks are the key attributes in designing a Bluetooth sensor system. The appropriate values of these features are contingent on the objectives of the intended application. It is worth mentioning that Bluetooth data collection systems, unlike GPS systems and cellular networks, require additional infrastructures (i.e., Bluetooth detectors) to be installed in transportation systems.

*5.3.2 Transportation-related applications of Bluetooth*

Since a decade ago, potential applications of the Bluetooth technology in traffic monitoring and management, as a noninvasive wireless data collection method that has no impact on the existing traffic conditions, have been investigated. A vehicle's travel time between two successive Bluetooth detectors is calculated by measuring the difference in time between two consecutive sensor stations that the vehicle has crossed from. Since Bluetooth detectors are capable of reading vehicles' MAC addresses within a specific range, the measured travel time is between two zones rather than two points. Hence, among multiple times that a vehicle's MAC address is scanned within each Bluetooth's detection zone, only the first-to-first or last-to-last detection needs to be matched in the travel time measurement for keeping consistency and reducing error (102). Considering the fact that the locations of detectors are already known, a vehicle's travel distance can be easily computed by finding the distance between those two successive stations that the travel time measurement was calculated. The space speed is then obtained by having the travel time and the travel distance. Fig. 6 displays the concept of estimating travel times and speeds using a Bluetooth sensor system. Recognizing travel times and speeds based on vehicles' trajectories gives room to establish other traffic management and traveler information applications such as creating time-dependent O-D matrixes over a time interval (103), estimating traffic distributions on alternative routes (104), and predicting dynamics of traffic volumes (105). Furthermore, the time duration that a Bluetooth-enabled vehicle spends to pass through the coverage area of a Bluetooth detector can be used as a quality-performance index of transportation systems. For example, in case of installing Bluetooth scanners at intersections, investigating the relationship between the link travel time and duration time leads to perceiving the level of service and delay at intersections (106).

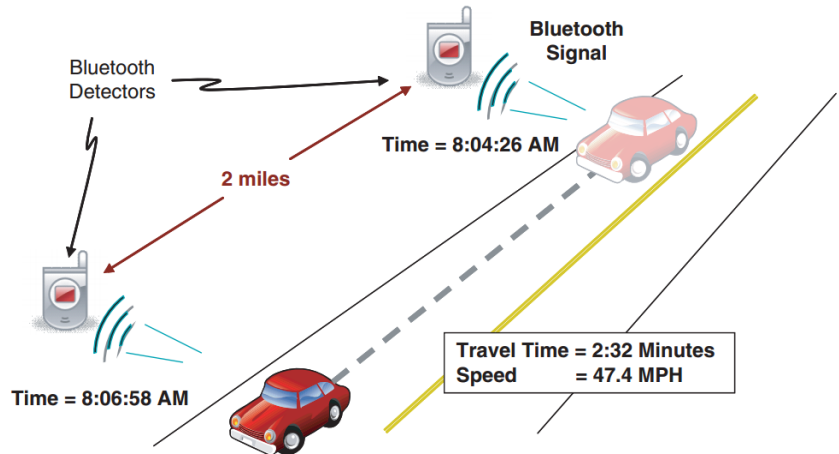

Fig. 6. Estimation of travel time and speed in a section using Bluetooth sensors.
(Source: Data collection of freeway travel time ground truth with Bluetooth sensors.
Transportation Research Record: Journal of the Transportation Research Board, 2010)

Bluetooth data are subjected to inherent sources of errors and noise that can be categorized into three groups: spatial error, temporal error, and sampling error (102). Spatial error refers to non-similarity of Bluetooth readers' detection zones that is dependent on amplitude of signals emitted by home appliances and cell phones. The second source of errors stems from variations in the detection time after a vehicle enters the detection zone of a Bluetooth reader since detectable characteristics of readers such as their signal strength and sensitivity are disparate. The third group points out to the possibility of having multiple Bluetooth devices in one vehicle, which results in duplicating one instance more than once in the database. Also, cyclists and pedestrians that move fast may be also counted as vehicles since Bluetooth readers are not able to distinguish whether the scanned devices are located inside the vehicles or other transportation modes. Thus, robust and effective filtering methods have been proposed to eliminate outliers and time intervals with either low number of observations or a large fluctuation in individual observations (107). The constructed error models and filtering methods can be evaluated using the travel times calculated based on inductive loop detectors and license plate reader data as ground truth.

### 5.4. Potential future research directions

A future research direction on prediction of traffic states using the FCD data is to incorporate other factors including (a) prevailing traffic conditions (e.g., percentage of heavy vehicles, local bus stations, and pedestrians flow), (b) roadway geometric conditions (e.g., grade and parking conditions), (c) weather conditions, and (d) traffic signal control settings. Moreover, the extent of uncertainty and reliability of many postulated distributions for traffic states such as travel time have not been taken into account. Consequently, in addition to the FCD data, other passive and active crowd sensing technologies can be employed so as to reduce the uncertainty in forecasting traffic states' distributions for large-scale traffic networks.

Notwithstanding the discoveries in the human mobility probability distribution, the rapid changes in the peoples' lifestyle, due to the growth of technologies and modern quality of thoughts, have been imposing changes in the human mobility patterns, which call for the continuous assessments on the human mobility. Thus, the movement data should be linked with their contextual spatial, environmental, and socio-demographic information to generate frameworks for a better understanding of human behavior. One particular characteristic of the human mobility is the choice of travel modes. In this sense, the transportation mode inference from GPS trajectories is limited to a few transport modes such as walk, bus, train, bike, and car. However, according to the FHWA, vehicles are categorized into 13 classes, where the passenger car is only one of them. Integrating spatiotemporal trajectory data with other sources of information provides the opportunity for advanced learning algorithms to extract additional movement features, which ultimately leads to classifying motorized vehicles into more subgroups.

Mobility studies that are constructed based on crowdsourced data convey several inevitable challenges such as randomness and a fixed sampling rate (81), which necessitate innovative strategies to relieve their effects. For instance, since users' personal information such as habits, social activities, and sickness can be inferred using only their location (108), their privacy and security need to be protected when users' location-based data are collected or implications of their mobility pattern analysis are published (109). Syntactic models of anonymity and differential privacy are two widely acceptable privacy models that target privacy-preserving data mining (PPDM) and privacy-preserving data publishing (PPDP). Yet a compromise between the risk of privacy violation and utility of data should be taken since privacy models such as noise addition techniques might directly impact on original data and learning algorithms (110).

## 6. Location-based social network

Today, social networking services have attracted millions of people around the world to share their thoughts, photos, locations, and videos. According to the Business Insider Intelligence report, approximately 20% of the total time spent on the internet is on the social media networks such as Facebook, Tweeter, Instagram, and Foursquare. In this context, an interesting question then arises: what type of hidden knowledge can be extracted from social network services for improving transportation systems?

Before proceeding to describe applications of social network services in transportation systems, the most related sort of social networks alongside its distinct attributes need to be introduced. Among a variety of social networking services, the Location-Based Social Network (LBSN) that generates users' spatiotemporal information is the most useful type of social media in transportation-related applications. The most distinct feature of an LBSN is enabling users to share location-embedded social contents that results in understanding users in terms of their location (111). A user can find and insert the location associated with the social media posts by means of location-acquisition technologies such as GPS that have been equipped in electronic devices. However, if accessing to location services has been permitted by users, the geographic location of the device that a user uses to post is automatically recorded. Foursquare, Twitter, and Flicker are the most widely used examples of LBSNs.

The service for sharing the location in LBSNs is categorized into three groups: Geo-tagged-media-based, point-location-based, and trajectory-based. In the first group, the location of shared media contents such as texts and photos is automatically labeled by devices or manually by users. Flicker, Panoramio, and Twitter are the representatives for such a service. Foursquare and Yelp are the social networks where people share their current location using the check-in feature as the point-location-based group. Evolving social networks are adding an option for recording users' GPS trajectories, which results in providing new information such as travel time, speed, and duration of stay in a location (112).

Data acquisition in social network services is typically processed through an Application Programming Interface (API). APIs are a set of protocols and programmatic access that enable developers to launch queries for accessing to both historical messages and real-time information. An API consists of several methods for filtering the desired contents matching a specified query. The API methods accepts various parameters for building a query, including a time period, a geographical region with GPS coordinates, specific keywords, users' name, or a combination of them (113). The API methods are often restricted to a specific rate limit for retrieving desired contents, which is a challenge specifically for real-time applications.

Users and locations are the major research topics in LBSNs although a strong correlation exists between the two (112, 114). Location-tagged contents indicate visiting locations of users. Connecting these locations sequentially with their timestamps leads to building a trajectory for each user. By tracing the location history of users, similarities in users' behaviors and interests are able to be extracted. Understanding users' activities and interests bring about several applications including friend recommendations upon mutual interests, finding local experts in a region, and community discovery (112). However, the location-perception topic is more associated to transport-domain applications as it gives room to mining users' trajectories and travel patterns. With mining users' travel sequences and processing the corresponding geo-tagged social media contents, a broad spectrum of transportation-related applications has been developed including travel recommendation systems, travel demand analysis, travel patterns and human mobility, urban planning, incident detection, emergency systems, and public transit, as depicted in Fig. 7.

## 6.1. Transportation-related applications of LBSNs

*Travel recommendation systems*: Using the LBSNs raw information and according to individual's preferences and constraints such as time and cost, a location recommender system provides the user either stand-alone locations such as a restaurant or a series of locations in the form of a travel route. Users' Geo-tagged social media and GPS trajectories are the typical data sources for providing the sequential location recommendation. For instance, geo-tagged photos generated by users in an LBSN such as Panoramia can be mined to provide a customized and automatic trip plan that contains three modules including destination discovery, discovering internal paths within a destination, and travel route suggestion plans. The modules optimize the best plan according to users' travel location, intended visiting time, and potential travel duration (115). Generating incomplete paths is the main weakness of geo-tagged photos since tourists and visitors are not necessarily taking and uploading photos in all locations they are visiting. Aggregating all photos taken in the same location and designing an algorithm to produce all possible travel routes by merging incomplete travel paths is a solution to alleviate the incomplete path issue. In (116), using Flicker geotagged photos, top ranking travel destinations and the best travel route between those destinations are detected and recommended while the main criteria for assessing the best route are the tourism popularity and the minimum distance to the destination.

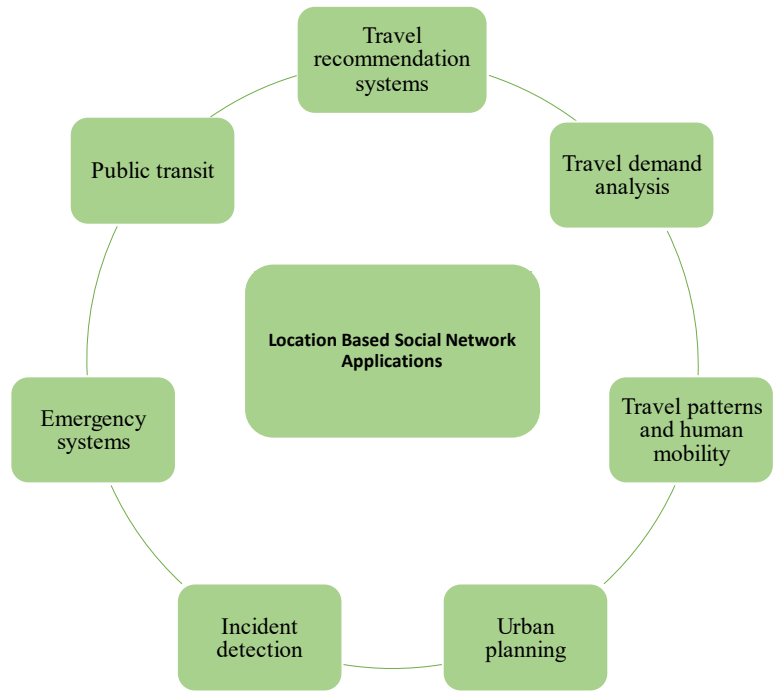

Fig. 7. Main applications of LBSNs in transportation systems.

*Travel demand analysis* is one of the active transportation research fields that models the mobility of people and vehicles to estimate travel behavior and travel demand for a specific future time window. Traditional 4-step modeling, as the most widely used model, forecasts the number of trips with a specific transportation mode made on each route between every pair of traffic zones. Foursquare check-in and geo-tagged Twitter data have been utilized to generate an origin-destination model, which is the second step in the 4-step modeling (117, 118). The results of studies verify the reliability of O-D matrixes created upon social media data as a cost-effective and non-time-consuming approach by comparing the proposed model with expensive and large-scale approaches such as household surveys (118, 119). In (118), for instance, millions of geo-tagged tweets in the Greater Los Angeles Area have been collected and mined to extract the O-D matrix. The proposed methodology consists of two steps: individual-based trajectory detection and place-based trip aggregation. Considering traffic analysis zones (TAZs) in the network, an individual OD-trip is defined when a user generates two consecutive tweets in different TAZs within a threshold time, which was set to 4 hours in this example. In the second step, the extracted trips for each pair of O-D are aggregated at different time windows such as an hourly or daily interval. Content analysis techniques can be applied into geo-tagged tweets to extract the distribution of travel mode choices in a specific area (120). Beyond trip-based models, accessing to large-scale and high-resolution individual level travel information has provided the chance to shift from aggregate-based models to individual level travel demand models such as activity-based models. These models reflect travelers' behavior attributes such as trip purpose, departure time, mode of transportation, activity location and duration, travel route, and traffic condition (121). A detailed discussion on how to obtain the mentioned individual travel information, as elements of the advanced travel demand framework, from social media data using data mining techniques is available in the reference (121).

*Travel patterns and human mobility*: Recently, a variety of research endeavor has sought to uncover underlying patterns of trips and human movement behavior from the social media data (122-124). After identifying the sequence of geo-location data for individuals, the most important characteristic of human mobility patterns, which is the displacement distribution such as power-law or exponential or a hybrid function, should be fitted according to the trajectory data collected through social media data. In the inter-urban movement analysis, the spatial interaction between cities can be determined using gravity models, in which the power-law distribution is a widely-used distance function (122). Spatial interactions between regions are rich information on understanding the spatial structure and community boundaries of a network. In intra-urban human mobility, the urban area is divided into a number of cells and a trip length is considered as the distance between centers of two consecutive cells that an individual has traveled. Trip purposes can be recognized from the type of check-in locations such as home, work, entertainment, and shopping. Accordingly, an activity is defined as a triple of the particular cell area, check-in time and venue. By finding a transition probability between two activities, an agent-based modeling can be deployed to reproduce the observed human mobility (125).

Furthermore, multi-day activity patterns of an individual can be inferred through Twitter messages contained Foursquare check-in data that show the type of activity (e.g., shopping and education) with the corresponding time and location (126). In spite of significant advantageous of social media data such as being large-scale, broad coverage and low cost, it accompanies with constraints including the lack of detailed description of activities, the absence of individuals' socio-demographic information, and inability to cover all activities since users are not sharing all sorts of activities (127). However, machine learning and pattern recognition approaches have been introduced to find both predefined and new patterns in people's travel behavior, despite existing of limitations in the geo-location data.

*Urban planning*: Geolocation social media data support the urban planning field and understanding of cities' dynamics, structures, and characters on a large scale (128). City planners can use social media data for introducing metrics that show interactions between people, understanding people's problem, and identifying potential solutions for improving cities (129). Mapping mobility, analyzing the design of physical spaces, and mapping sociodemographic status are examples of indices that are potential to be captured by the social media and deployed in improving the quality of life in cities. The structure of local urban areas and neighborhoods can be inferred by clustering nearby locations from check-in Foursquare data based on two criteria: spatial proximity and social proximity. While the former criterion measures the closeness of venues to each other, the latter one is the social similarity between each pair of venues based on the number of times a venue is visited (128).

*Monitoring traffic condition and traffic incidents*: A powerful traffic monitoring and incident detection system needs accessibility to complementary sources with a high coverage area such as social media data. On the contrary to existing data sources such as inductive-loop detectors and traffic cameras, social media data are not limited by sparse coverage, which in turn can be considered as a ubiquitous and complementary data source in traffic information systems. Among various popular social networks, Twitter has been received more attention for extracting traffic conditions due to using short messages. The limitation on the tweet length, which is 140 characters per tweet, forces people to communicate in a timely and effective fashion. To discriminate traffic-related tweets from non-traffic-related tweets, several studies have exploited text mining and machine learning techniques to automatically extract useful and meaningful information from the unstructured and irregular text of tweets. Bag-of-words representation is a typical approach for mapping tweet texts into numerical feature vectors before feeding them into a supervised learning algorithm for the classification task (130, 131). However, it creates a high-dimensional sparse matrix, which calls for compressing the numerical vector space into only a set of traffic-related keywords to limit the number of features (132). Although such a strategy engenders a much-lower-dimensional matrix, the immediate critique of using a pre-defined set of keywords as features is that the vocabulary may not include all important traffic-related keywords. Furthermore, bag-of-words representation completely ignores the temporal order of words, which leads to missing the semantic and syntactic relationship between words in a tweet. One way to address these shortcomings is to utilize the semantic similarity between the words in a tweet using word embeddings tools (133, 134). A word embedding model maps millions of words to numerical feature vectors in such a way that the similar words tend to be closer to each other in vector space. However, the most efficient way for detecting traffic-related tweets without needing a pre-defined set of keywords is to deploy deep learning architectures such as convolutional and recurrent neural networks (135). These models, which are very popular in semantic analysis and text categorization, have high capability to capture both local features of words as well as global and temporal sentence semantics (136).

*Emergency systems*: A large and growing body of literature has investigated the role of social media in emergency and disaster management systems since the past decade. Imran et al. (113) reviewed the cutting-edge computational methods for processing and converting the raw information of social media into meaningful and actionable insights for managing emergency events such as natural disasters. The study presented methods on social media data acquisition and preprocessing, detection and tracking emergency events, clustering and classification of messages for extracting and aggregating useful information, as well as applications of semantic technologies in a crisis situation. Twitter is the most widespread social media platform used in the literature of emergency management due to its distinct features such as the ability of users to post tweets on a real-time basis during emergency cases and providing a higher number of queries and data. In (137), as a well-structured example for the application of Twitter in the real-time emergency event detection, the target event is first detected by classifying tweets into positive/negative labels using the support vector machine algorithm when features include keywords in a tweet, the number of words in a tweet, and the word before and after the query word. Afterward, the location and trajectory of the target event are estimated using probabilistic spatiotemporal models such as Kalman and particle filtering.

*Public Transit*: In the public transportation research area, social media data have been utilized to assess transit systems operation according to public opinion and attitudes that can be inferred by a sentiment analysis (121). The strategies that public transport agencies have deployed to arrange their social media programs, associated goals, and measurements for assessing their programs have been studied in the reference (138). Results of statistical analysis, that was applied to information collected through online surveys filled out by top transit agencies, indicate that social media can be used to proliferate transit environmental benefits, safety, and livability improvements. A comprehensive literature review on the role of social media in managing the public transit has been conducted in the reference (139).

*6.2. Potential future research directions*

Spatial-temporal sparsity is the main issue in the LBSNs information as users do not necessarily share geo-tagged media contents during their travels, reflecting in incomplete trajectories. One way to deal with data sets that contain missing values is to apply advanced strategies such as collaborative filtering. Then, missing values are imputed with a set of plausible values that represent the uncertainty about the correct values, which ultimately results in transferring the sparse original dataset to a dense one. Applying the graphical inference models, such as Bayesian network and Markov random field, in the dense geo-tagged tweets or check-in records results in a more reliable framework for individuals' activities and mobility behaviors.

As mentioned, low-cost and large-scale location-embedded social media data have considerable potentials in advanced travel demand modeling and long-term urban planning goals. However, an activity-based travel demand modeler might encounter a couple of challenges for extracting the information regarding in-home activities, different types of daily activities, and future activities for all individuals from their limited shared social media contents (121). Other than deploying advanced learning frameworks (e.g., recurrent neural networks and word embedding models) to extract semantic and syntactic information on social media contents, enriching social media data by adding other sources (e.g., GIS-based land use information and GPS-based trajectories) clarifies unknown details in travel demand models including the number of daily trips with different purposes, visiting locations, travel routes, and transport modes. Like all crowd sensing technologies, privacy issues need to be taken into account so that the personal information is untraceable and irretrievable from outputs of travel demand models.

## 7. Smart card and automated passenger counter

Smart card Automated Fare Collection (AFC) and Automated Passenger Counter (APC) are technologies for collecting public transit data in order to both describe the spatial-temporal patterns of passengers' behavioral and evaluate transit facilities. These two sources of transit data can be supported by openly General Transit Feed Specification (GTFS) file, which contains publicly-accessible public transportation scheduled operations (e.g., daily service patterns) and network geometry information (e.g., stop locations) that have been published by transit agencies (140). Although APC has been particularly designed for counting passengers in and out of public transport modes (e.g., bus and subway), the main purpose of a smart card is to collect revenue. In the following sections, we briefly present how AFC and APC work and what kinds of raw data they can collect. Then, we review a vast majority of studies that have disclosed applications of these emerging technologies on transit system planning and operations. Compared to relatively few historical studies in the area of APC, a great deal of previous research into public transit data has focused more on smart card systems on account of its ubiquity and ability to collect a large-scale spatiotemporal data. Thus, we mainly elaborate the applications of smart card systems even although the data provided by both systems are interchangeable and complementary in several applications.

*7.1. Operational mechanisms of smart card AFC and APC*

Smart card is a device designed for storing data and equipped with an embedded chip on which information is stored. Smart card technologies can be divided into two groups: contact and contactless card. In the contact type, the embedded chip is not covered with plastic, which the card needs to have a direct physical contact to be connected with a reader. However, in the contactless smart card, the card contains not only the chip, which is completely embedded into the card, but also an antenna. The latter component enables the card to have a remote communication with radio frequency identification (RFID) devices through high-frequency waves (141). By bringing the card close to a reader, the chip is powered through the electromagnetic field of the reader. Then, wireless communication, based on high-frequency waves, is established to transfer data between the card and the reader. As the contactless cards are more secure, more transit agencies in the US have used contactless smart cards for AFC systems since the 1990s (142). By moving a smart card close to the reader of a public transport station, a set of typical data are being collected: validation status of the smart card (expiration date, compatibility with type of service), status of transaction (boarding acceptance/refusal, transfer), card ID, route ID/direction, stop/bus ID, date and time of operation (141). All data stored in a reader are transferred to the central server on a regular basis.

An APC is a device installed on transit vehicles such as a bus and a light train to count the passengers who get off and get in the transit vehicle at each station. The goal is to improve the accuracy of counting ridership at the disaggregate level, which is appropriate for service planning and schedule routes. APCs work based on two types of electronic units: infrared beams and video cameras. The beams of infrared lights shine down to the person crossing the stairwells, where the laser is located mounted on the ceiling (143). The way that a person breaks the beam determines boarding or alighting activities. Video counting system is the second type of units used in APCs, which increases the accuracy of counting by 98% in comparison to the manual counting collected by a checker or on-board surveys (144). A video-based-people counter sums up how many passengers get on/off. A definite merit of video camera systems is the ability to verify counting process by watching the video back. In addition to counting people on and off the vehicle, the dwell time at each station and the passengers flow around stations are other important features recorded by APCs. Integrating APC systems with on-board GPS receivers leads to detecting the location of counted passengers. In a similar manner to smart card systems, the stored data are wirelessly transferred to a central server.

## 7.2. Transportation-related applications of smart card systems

In addition to the aforementioned smart card data, boarding date/time/location are other principal data provided by smart cards for further analysis in understanding public transit operations and users' motilities. GTFS and APC data, as complementary sources, have also been combined with the smart card data source in order to enhance the ability of realizing various aspects of transit networks. The location of bus/subway stations can be obtained directly from the database of operating transit agencies (i.e., GTFS). For those transactions made inside the bus, the boarding location is derived by matching database from other positioning systems such as GPS receivers equipped in the bus fleet. Unfortunately, alighting information for tap-in-only stations cannot be accessed through smart card systems, yet a group of research scientists has developed models for forecasting alighting locations (145). Having individual boarding and alighting location data, the route profile load can be computed as well. Also, the absence of socio-demographic information of cardholders in the smart card data calls for other complementary sources such as household surveys to enrich smart card data before conducting strategic planning (146). In 2011, Pelletier et al. (141) reviewed the literature on the use of smart card data in public transit systems in their solid survey paper. They classified studies in this context into three folds: 1) long-term planning including transit demand modeling and behavior analysis, 2) service and schedule adjustment, 3) operational metrics. Accordingly, in this paper, we are mainly reviewing studies

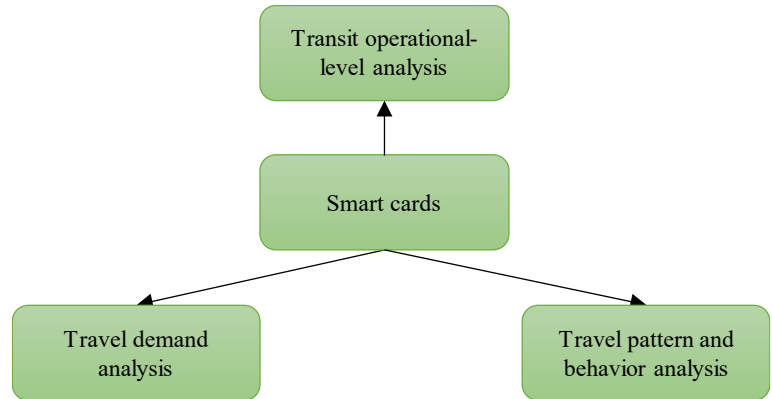

Fig. 8. Main applications of smart card in transportation systems.

that have been conducted after 2011, and refer the remaining literature to the work of Pelletier et al. Fig. 8 displays three main groups of smart card applications in transportation fields. *Transit operational-level analysis*: Summarizing and visualizing the main characteristics of public transit data are the first steps to take benefits of such large-scale data. Basic statistic tools such as Exploratory Data Analysis (EDA) and probability distribution models empower practitioners to transfer a large volume of data into useful information on transit services and operations. Distribution of minimum-distance access to transit stations, daily smart card usage frequency and bus/subway stations usage frequency are examples of hidden knowledge in public transport services that can be easily revealed through EDA as well as descriptive and numerical statistics (147). Furthermore, in order to enhance users' satisfaction and transit management systems, performance indicators on both transit operation (e.g., schedule adherence, vehicle-kilometers/hours, travel time and speed, and headway variance) as well as transit usage (e.g., passenger kilometers/hours, and the number of boarding and alighting passengers) need to be calculated for every individual run segment, bus stop, route, and day (148)

*Travel pattern and behavior analysis*: Understanding passengers' travel behavior leads to a variety of applications. First of all, authorities are able to provide oriented services and information for identifiable classes of similar behaviors (149). Assessing the performance of a transit network, detecting irregularities, forecasting demand, service adjustments, analysis of minimum-access distribution to transit stations, are other application examples of passengers' behavioral analysis (150). Data mining techniques can be also applied to individual level trip chains (i.e., a series of trips made by an individual in each day) to extract passengers' travel pattern and travel regularity (151). Both high-likelihood travel patterns of an individual rider and similar patterns among all transit riders, which is also known as the aggregated level, have been recognized using data mining algorithms based on riders' temporal and spatial characteristics such as the number of travel days, the number of similar first boarding times, and the number of similar stop-location-ID sequences. After identifying passengers' travel pattern based on specific attributes (e.g., boarding information), the relationship between passengers' socioeconomic characteristics and travel patterns can be discovered by assigning passengers to their residential area retrieved from boarding information. Income and total living area per household are exemplified as the socioeconomic data (150). In addition to the passenger aggregation, travel pattern analysis has been carried out in the stop aggregation level. Based on the premise that an activity is associated with a specific location rather than an individual stop, those transit stops that can be used alternatively by passengers for doing a specific activity are aggregated to the same travel pattern according to geographical proximity between individual stops or similar names among stops (152). Classifying users' travel behavior at the trip level is another implication of mining smart card transactions. Transfer journeys are able to be distinguished from single journeys according to the boarding information (153).

*Travel demand analysis*: Transit O-D matrix is valuable information for not only making a balance between supply and demand in transit networks but a better understanding of individuals' travel patterns. A trip in the O-D matrix is defined as a movement that starts from an origin location and terminates in a destination point. For transit stations without a tap-out facility, a user's trip, that contains segment(s) between two consecutive boarding and alighting locations, is constructed by estimating the alighting points while the boarding location is already obtained through smart card transactions (154). Several assumptions have made in literature for predicting the alighting station of a segment and in turn the trip destination. Two primary assumptions include: the origin of the next trip is the destination of the previous trip and the origin of the first daily trip is the destination of the last one (155). Considering such assumptions, the nearest station to the next boarding bus stop within a walking distance threshold or an allowable transfer time (e.g., 400 meters or 5 minutes) is chosen as the alighting station of the previous trip. Discovering the alighting positions by following the location and time of the next boarding point is the main idea to infer the final destination of trip and as its consequence to create the transit O-D matrix (154). Recently, the transit systems equipped with a tap-out device payment have provided alighting points data, which give room to assessing the previous methods and assumptions in literature (156).

Trip purpose or type of activity is another salient feature in the travel demand modeling that can be derived from users' information (e.g., fare card type), temporal information (e.g., transaction time), and spatial information (e.g., land use type of boarding/alighting locations) (157). Applying a rule-based classification on the mentioned features leads to building a trip-purpose inference model. For instance, suppose two transactions are observed in a user's metro smart card, the first one in the early morning and the second transaction in the late afternoon. If the boarding and alighting points are residential and downtown areas, respectively, the trip purpose is inferred as a work-related activity.

### 7.3. Potential future research directions

Moving towards modern and smart forms of ticketing systems, the smart card AFC technology has become acceptable as an efficient, profitable, and convenient payment system in public transport networks for both transit agencies and customers. As a result, millions of transactions are completed in each operating day, which ends up with generating vast amounts of smart card records in exponential growth. Beyond the dire need for complex distributed storage and computing platforms such as Hadoop, advanced preprocessing techniques are essential to organizing such massive data. Accordingly, before using learning algorithms for extracting travel patterns, improved data validation and data cleansing methods need to be applied to smart card data to ensure that high-quality data are fed into learning algorithms. However, a majority of studies that use smart card data for developing transport-domain applications have focused on only the analytical part without paying heed to the fact that the final implications of travel patterns and demand analyses are primary contingent on to what extent the smart card data are validated, clean, accurate, complete, fitted, and uniform.

One of the main challenges in entry-only AFC systems is the scarcity of alighting information because no facility exists to record the smart card fare data when passengers alight and exit the station. The fundamental assumptions, mentioned in the previous sections, for estimating the alighting location might be invalid in many circumstances. In case that a passenger's journey contains an intermediate segment with a transport mode other than a bus or a train, the assumption that the most likely destination of a passenger trip is near to the origin of the next trip does not hold anymore. Also, passengers' mode choices towards the work might be different from returning home; where, for instance, a passenger uses buses in the early morning while they can carpool with family on the way back to the home. Thus, the assumption that the origin of the first daily trip is near to the destination of the last one might be invalid. To overcome the issue, many research endeavors have been conducted to validate only the destination inference based on external data sources such as O-D trip surveys or integration of AFC with other facilities such as Automated Vehicle Location technologies (158). In this context, a future research direction for inferring a high percentage of the journey destination, when the assumptions are broken and no external infrastructure is available, is to incorporate other spatiotemporal sources such as related GSM data so as to directly extract the alighting information from the fused data.

## 8. Environmental Data

Meteorological data and air quality data are two main groups of the environmental data (7). The first group consists of atmospheric information including humidity, temperature, barometer pressure, wind speed, evaporation, precipitation, snowfall, types of weather, etc. The related public websites (e.g., climate.gov and weather.gov) and agencies are the primary sources for collecting the meteorological data. National Climatic Data Center (NCDC) is one of the three data centers in National Center for Environmental Information (NCEI), which is a trusted authority for hosting and providing the weather data and information across the United States (159). Land-based meteorological data over specified date range and location is one of the main products in NCDC. The data are collected through instruments located in the field.

Air quality data indicate the concentration of various pollutants in the air such as carbon monoxide (CO), lead (Pb), ground-level ozone ($O_3$), nitrogen dioxide ($NO_2$), sulfur dioxide ($SO_2$), and particulate matter ($PM_{10}$ and $PM_{2.5}$). Installing permanent monitoring stations throughout the cities is the most common way to measure and forecast how much the air is polluted based on the above contaminants, where the degree of pollution is reported as Air Quality Index (AQI). The AQI ranges from 0 to 500, in which the lower values imply the good air quality with low-to-moderate health concerns, but the higher values pose the entire population in a hazardous condition. Daily air quality data for a specific location from the corresponding monitor are available by the US Environmental Protection Agency (EPA) (160). Transportation sector has become a major cause of air pollution by burning fossil fuels alongside with many other factors including manufacturing activities, production of electricity, etc. Hence, with regard to different causes of air pollution and location-to-location variation of the traffic flow and the land use, the measurements of a ground-based station only relate to its adjacent area and cannot be expanded throughout the city (7). Such a limitation besides the high expenditure of installing and maintaining monitor stations call for alternative strategies. The techniques for estimating air quality and the benefits of understanding air quality distribution throughout cities are summarized in Fig. 9.

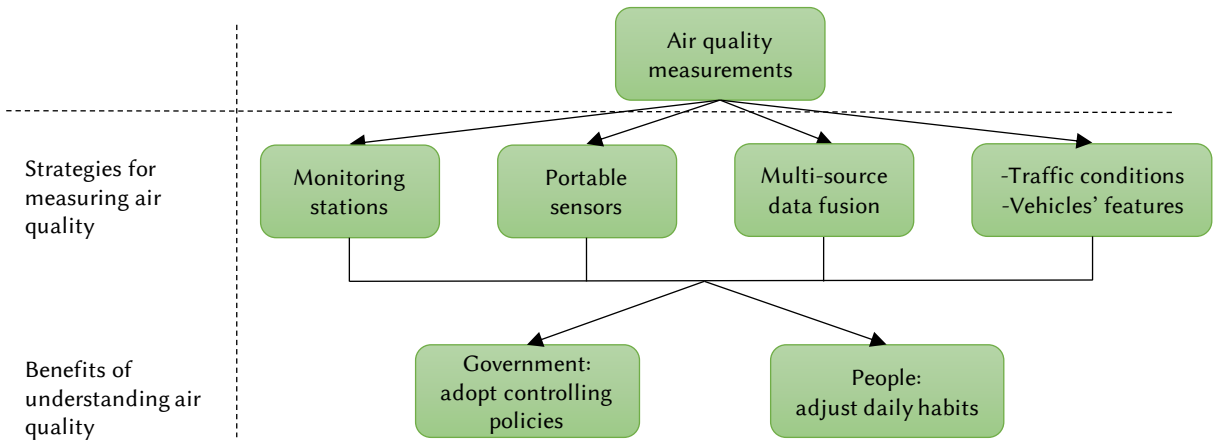

Fig. 9. Structure of air-quality-measurement techniques and the associated benefits.

One way to measure air pollutants in a wider area more than the point locations of monitoring stations is to use crowdsourcing attribute of mobile instrumentations and sensing technologies. Environmental data can be sensed with high resolution by installing portable sensors on bikes' wheels, or putting a mobile sensing box equipped with air pollution sensors inside vehicles. In accordance with the mobile sensing box installed on vehicles, fine-grained air quality with respect to each type of pollutants can be measured and monitored dynamically (161). The captured air pollution data are tagged with the time and location with the help of either the GPS receiver and a cellular modem attached to the mobile sensing box, or the on-board location-acquisition technologies such GPS-equipped devices.

Thus far, several studies have attempted to estimate traffic emission and fuel consumption upon traffic flow characteristics, vehicle-related features, and drivers' behaviors. In the macroscopic level, emissions are calculated based on the average speed/travel distance of vehicles and the emission factor, which is a function of pollutant type, road class, vehicle category, and fuel type (162). On the other hand, microscopic models have utilized instantaneous vehicles' speeds/accelerations and drivers' route-choice decisions in computing emission measurements and fuel consumption (163, 164). Furthermore, the emission rate has been modeled by incorporating other factors including roadway conditions (e.g., length/slope of roads) (165) and climate conditions such as the weather temperature (166). Currently, MOVES is the latest model for emission estimation developed by EPA. The model is able to predict air pollution emissions produced by cars, trucks, and non-highway mobile sources based on settings specified by the user including vehicle types, time periods, geographical areas, pollutants, vehicle operating characteristics, and road types (167). It is worth mentioning that such traffic-based emission models only consider the part of air pollution caused by vehicles.

Applying machine-learning techniques on multi-source air quality data is another alternative for estimating fine-grained air quality distributions for the entire city, rather than only the areas adjacent to monitor stations. Such an inference can be derived from not only the monitoring stations as the main and ground-truth source (168) but a combination of other sources ranging from meteorological variables and land use information to traffic flow characteristics based on vehicles' trajectories and road network features (169). After gleaning spatiotemporal features from the observed data, the air quality index in different regions of a network can be inferred and clustered using machine learning techniques such as self-organizing-map neural network (168, 170) and conditional random field (169).

Understanding present and future air quality distributions throughout cities serve a precious reference for government agencies to adopt new policies for controlling and improving air quality such as imposing different road tax rates for various levels of fuel consumption (171) or granting financial support to industrial owners who move towards emissions reduction (168). Air quality information has also a profound impact on people's daily habits such as using face masks appropriately and choosing the best time/location of different activities, which are effective ways to keeping their health and preventing medical treatment.

### 8.1. Potential future research directions

Revealing the citywide and real-time air quality information to drivers and people affect their daily activities and travel behavior. Given the dynamic air quality information throughout a city, drivers' behavior in the route and mode choice decisions can be analyzed. Many users may prefer to commute through less polluted roads although their journey time may increase. Today, due to the complexity of roadway geometry and dynamic changes in traffic conditions, drivers are directed towards their destination through an optimal route by means of in-vehicle or smartphone-based navigation systems so as to avoid being lost and save their journey cost. Accordingly, the navigation system can be configured with the fine-grained air quality information to guide drivers along the cleanest route as an alternative to the fastest route.

A related gap in the air-quality measurement literature can be filled by incorporating the microscopic and macroscopic vehicle-based emission models into Big-Data frameworks of fine-grained air quality estimation. For instance, in the co-training framework proposed by (169), the proposed emission models, that are typically the functions of vehicles' trajectory features such as speed and acceleration, can be fed into a fusion framework instead of directly feeding speed-related features into classifiers. Moreover, the fuel consumption and emission rate impart the prior knowledge on air quality conditions in the data fusion framework.

## 9. Data-fusion architectures

Each of the above-mentioned sensors generates massive traffic data that can be deployed for predicting and managing traffic states, including travel time, vehicular speed, vehicle classification, traffic density, and traffic volume. However, measurement errors, uncertainty in an individual source, inability to collect all kinds of data, incapability of a sensor to detect all types of objects, sparseness and limitation to a specific point or a small area are the major weaknesses of deploying only a single data source. Fortunately, with the advent of technologies and ability to collect data from multiple sources, data fusion techniques are applied to independent but complementary data sources in order to obtain a better understanding of mobility behavior and travel patterns that could not be attained from a single data source. The main goal of data fusion algorithms in the smart transportation is to develop an improved control system that fuses the knowledge learned from a set of heterogeneous data sources. Data fusion frameworks aim to provide sufficient information for estimating traffic parameters and making appropriate decisions that ensure users' safety and an efficient traffic stream (172).

In practice, data fusion topologies are distinguished from each other according to the selected criterion for fusing data. The selected criteria are defined as follows (173): 1) The relations between data sources, where the types of relationships are complementary, redundant, and cooperative. 2) The input/output data types, which are subdivided into different combinations of data/feature/decision in versus data/feature/decision out. 3) The abstraction levels of the deployed data in fusing algorithms, where the raw data, features, and decisions are three levels of abstraction that are transformed to the more accurate levels by feeding into fusion algorithms. 4) The high-level technique for implementing the fusion process, which consists of four types of fusion architectures: (a) centralized architecture: information from all data sources are transmitted to the central processor that performs the data fusion process, (b) decentralized architecture: the complete data fusion process and analysis are conducted in each data source node autonomously. Then, each node provides its own statistical model, and only the information is exchanged between the nodes. This architecture can be also viewed as the stage-based fusion where each data source is used at different stages of the fusion framework, (c) distributed architecture: first, each source node processes and analyzes its own information to build a separate model; and then, the constructed models, which are the outputs of each data source, are fed into the fusion node as inputs (or new representation of the original features) for generating a more accurate model, (d) hierarchical architecture: the decentralized and distributed architectures are combined, where the fusion process is performed at various levels of the hierarchy. Fig. 10 represents the differences among the described architectures. Notwithstanding that many statistical, probabilistic and artificial intelligent algorithms can be used in data fusion architectures, the most widely used fusion algorithms in transport domains include, but not limited to, weighted average methods, Kalman filtering, particle filtering, Monte Carlo techniques, Bayesian inference, Dempster-Shafer evidential theory, and artificial neural network (172). The choice of an appropriate algorithm is highly dependent on the subject application and the features of available data.

### 9.1. Transport-related applications of data-fusion techniques

Many traffic engineering problems and applications that have been explained throughout this paper can be also addressed by fusing those heterogeneous sources of information. In a seminal and survey study of data fusion applications in the smart transportation, El Faouzi et al. (172) summarized potential uses of data fusion algorithms in several functions of ITS, including advanced traveler information systems, automatic incident detection, advanced driver assistance, network control, crash analysis and prevention, traffic demand estimation, traffic forecasting and traffic monitoring, and accurate position estimation. Here, we review a set of notable examples of data fusion applications that have utilized a combination of traffic data sources. Examples have been selected in such a way to cover all mentioned traffic-related data sources.

In a study conducted by Choi et al. (174), data generated by six loop detectors and eight GPS-equipped probe vehicles were fed into an information fusion framework to calculate the link travel time. At first stages, the one-minute travel time calculated by each loop detector was extended to a five-minute travel time, using the voting technique, where the five-minute period is the adopted interval of information update and analysis. On the other hand, the five-minute travel time is also computed based on the GPS data of each probe, and then fused based on the fuzzy regression. As a final point, the outcomes of the two sources are integrated based on the Bayesian pooling technique to construct a more accurate representative of the link travel time in comparison with ordinary arithmetic mean of both traffic sources. Bachmann et al. (175) compared the efficacy of seven data fusion techniques for traffic speed estimation. Data were collected based on loop detectors and Bluetooth-based probe vehicles by means of a microsimulation model. The microsimulation package provides the opportunity to not only examine the impact of probe vehicle sample sizes on the model's accuracy but determine the ground truth traffic conditions, which is defined as the exact average link speed of all vehicles. Note that deriving such ground truth is almost impossible in the real world. Simple convex combination, Bar-Shalom/campo combination, measurement fusion Kalman filter, Single-Constraint-At-A-Time Kalman filter, Ordered Weighted Averaging (OWA), fuzzy integral, and neural network were utilized as the data fusion algorithms, where all was executed in the distributed architecture as shown in Fig. 10. The results indicated that, except OWA that was deprived of statistical basis, all data fusion techniques outperform the speed estimation computed by only loop detector data or probe vehicle data.

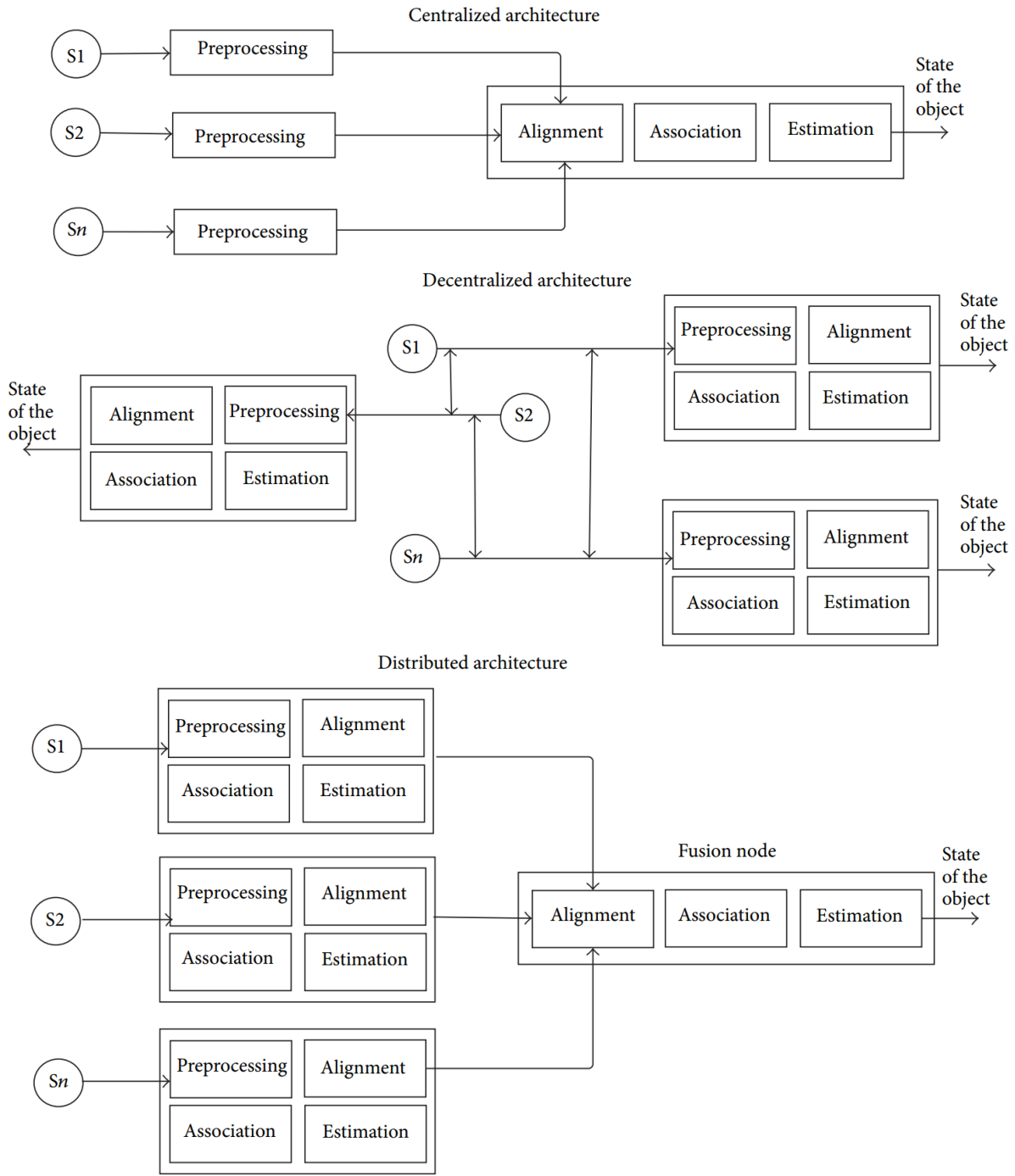

Fig. 10. Data fusion topologies based on the level that the fusion process is performed. (Source: A review of data fusion techniques. *The Scientific World Journal* 2013)

In an investigation of data fusion capability into the automatic vehicle guidance, Munz et. al (176) introduced a centralized architecture fusion system for devising a reliable vehicles' detection and tracking system, as the main component of the driver assistant system. They claimed that their proposed framework is also applicable to any types of sensors combination regardless of what kind of object they can detect and classify. To accomplish the task, the Joint Integrated Probabilistic Data Association algorithm and its extension based on the Dempster-Shafer Theory (DST) were combined to describe both sensors' properties and their environmental aspects. The postulated fusion framework was applied to the real-world data from the vehicle equipped with a sensor combination of a video camera and a laser scanner, which rendered a robust multi-class detection and tracking system.

The inference of long-term travelers' behavior from the smart card data can be improved when the absent behavioral attributes are supplemented using other sources. In (177), a data fusion framework was developed to estimate the trip purpose of each record in smart card data using the naïve Bayes classifier method. In their framework, the naïve Bayes probability is the multiplication of the prior probability of the trip purpose and the conditional probability of the trip attributes given the trip purpose. Both probabilities were derived from the person trip survey data by conjecturing that these two probabilities are the same as the corresponding probabilities in smart card data. Accordingly, the naïve Bayes classifier is able to both predict the trip purpose of each smart-card data item given the trip attributes (e.g., the boarding and alighting locations and time) and then reconstruct the smart card database with the trip purpose as a new behavioral feature. Another example data fusion application is the provision of a visual fusion platform to examine users' behavioral changes in a complex transportation network by fusing smart card and social media data (178). Furthermore, the created visual fusion facilitates tracking the occurrence of the unexpected phenomena, their spread throughout the network, as well as their causes and effects on the transportation system.

Usage of the LBSN data in data fusion frameworks was investigated by Pan et al. (96). Using a decentralized data fusion architecture, first, a traffic anomaly is captured and defined as the sub-graph of a network that its GPS-based drivers' trajectories are not adapted to normal patterns. Afterward, in order to ameliorate the accuracy of the anomaly detection process, the tweets posted in the area of the detected anomalous sub-graph are retrieved and mined to discover the representative terms such as "sport" and "accident" that imply the occurrence of an unusual event. Finally, as an example of the data fusion concept based on the environmental data, Zheng et al. (169) developed a co-training framework that infers the fine-grained air quality by fusing five different datasets: air quality monitoring stations, meteorological data, taxi trajectories, Point of Interest (POI) locations, and road networks. Since the air quality varies at different time and places, both spatial and temporal dependencies of the air quality need to be modeled. Thus, the datasets related to the spatial features (e.g., roadway geometry) and the temporarily-related datasets (e.g., traffic volume data) are fed into two separated learning algorithms to measure the spatial and temporal dependencies of the air quality, respectively. After training the spatial and temporal classifiers, air quality index is measured by the product of the scores obtained through the two classifiers.

## 10. Smart transportation: current challenges and solutions

Thus far, we elaborate the ways that multi-modal data sources and technologies have been deployed to build advanced models in a wide range of transport domains. However, data sensing and management, and analytic models (i.e., the leading steps in the smart city and smart transportation) have been confronting some common challenges. The purpose of this section is to warn the readers that although deploying and mining the aforementioned data sources bring many benefits, it is imperative to also pay heed to the key challenges accompanied with data sensing, acquisition, management, and analytical processes. Here, we only introduce the obstacles and provide references for more details. It should be noted that many issues that are introducing in this section still needs more advanced strategies, which in turn can be thought of as future research directions in the smart transportation.

In spite of the fact that many types of fixed-point traffic sensors have been installed throughout transportation networks, they have not covered all segments of the network. Installing the stationary sensing infrastructures in the majority portions of the city incurs large expenditures and redundant data, which is not a practical and intelligent approach. One way to meet this challenge, as mentioned, is to leverage probe vehicles and people as movable sensors by tracking their mobile phones, commuting smart cards, and social media activities. Although moving sensors support to realize the mobility behavior and travel patterns, on the opposite side, they produce new issues such privacy protection, non-uniform distributed sensors, and generating unstructured data (7). Regarding the invasion of privacy, collecting the individuals' GPS trajectories discloses the person's private affairs such as their visited locations and daily routines. Also, the data generated by moving sensors possess inherent issues. One is the sparsity and non-uniformity that originates from the fact that people are not necessarily turning on their device's location service to sharing information, which results in the lack of the sufficient movement data in some regions. Another type of issue, unlike the sources designed specifically for traffic management, is that the data produced by people are coming in a variety of formats such as texts and images, which are not only unstructured with a variable rate of resolution but require an initial inspection to assure they are traffic-related data.

Another existence challenge is that traffic data are received from various heterogeneous sources even for solving a specific issue. Data sources for constructing origin-destination matrix, for example, involves household interviews, mobile cellular networks, and GPS trajectories. Thus, while a majority of existing data mining algorithms analyze data from one source (e.g., computer vision with images and natural language processing with texts), advanced machine learning and data-fusion algorithms are indispensable to exploit mutual knowledge from multi-source and heterogeneous data. Furthermore, in many transportation scenarios such as an adaptive traffic signal control and in-vehicle collision systems, it is imperative to respond on a real-time basis. This calls for not only cutting-edge data management systems to organize the multiple-source traffic but advanced data fusion and analytical tools to promptly make the optimal decision in responding to dynamic changes in traffic conditions.

The similarity between the phases in the smart transportation framework and stages in the Big Data phenomenon (i.e., data generation, data sensing, data management, and data analysis (179)) hints at considering the smart transportation as an application of Big Data. As a consequence, in addition to what were explained in the previous paragraphs, the smart transportation involves challenges in Big Data, which are enumerated as: heterogeneous and inconsistent data, large scale of data, visualization raw and analyzed data, redundancy reduction and data compression, and safety and confidentiality of data. A broad spectrum of studies in literature and other ongoing research has been seeking to address these challenges. However, as scrutinizing the methods for resolving the mentioned issues is out of this article's scope, further details on Big Data problems are referred to the references (179-181).

## 11. Conclusion

Researchers and authorities have always been looking for smart solutions for resolving endless challenges caused by transportation activities. A cost-effective approach is to optimize the transportation network by mining the hidden knowledge in traffic-related data that are sensed through Information and Communication Technology infrastructures. In addition to the traffic flow sensors that have particularly been designed for recording vehicles' presence and passage, other existing technologies that have the potential to record people's mobility can be leveraged for improving the transportation network performance. People actively participate in the process of recording data about different aspects of their life using technologies such as GPS systems, cellular networks, social media services, and smart card AFCs. Such diverse and advanced sensing technologies deliver a large-scale data regarding mobility behavior of people and vehicles, which can be fed into the advanced learning algorithms and data analytic models so as to bring a variety of advantages, including modern traffic management and control strategies, robust vehicle detection systems, mining mobility patterns, travel demand analysis, incident detection, air quality measurement, urban planning, and emergency systems. The acquisition, management, and mining of traffic-related data constitute the core components of the smart transportation.

The aim of the present survey paper was to identify the transport-domain application of widely-used-traffic data sources. For each data source, we not only explained the operational mechanism of the technology used for generating and extracting data but also divided all related applications into major groups and indicated a number of possible future research directions. Furthermore, data fusion architectures for integrating these independent but complementary traffic data sources were investigated. The key strength of this study is the provision of an exhaustive guideline for readers to perceive the potential applications of multiple traffic data sources, which in turn assist them to comprehend the existing literature very promptly and brainstorm the future work. We also succinctly discussed on the current issues and the corresponding solutions of the smart transportation.